\documentclass[useAMS,usenatbib,referee]{mn2e}
\usepackage{graphicx}
\usepackage{natbib}
\usepackage{times}
\usepackage{float}
\usepackage{epsfig}
\usepackage[utf8x]{inputenc}
\usepackage{amsmath}

\newcommand{\cx}[1] {\cal{#1}}

\title[Twisted disc formed after a tidal disruption event]{
The dynamics of twisted disc formed after  the tidal disruption  of a star by a rotating black 
hole}
\author[P. B. Ivanov, V. V. Zhuravlev and  J. C. B. Papaloizou ]{P.B. Ivanov$^{1,3}$\thanks{E-mail: pbi20@cam.ac.uk (PBI)},
V. V. Zhuravlev$^{2}$\thanks{E-mail: v.jouravlev@gmail.com  (VVZ)}
 and J. C. B. Papaloizou $^{3}$\thanks{E-mail: J.C.B.Papaloizou@damtp.cam.ac.uk (JCBP)},\\
$^{1}$Astro Space Centre, P.N. Lebedev Physical Institute, 4/32
Profsoyuznaya Street, Moscow, 117810, Russia\\$^{2}$
Sternberg Astronomical Institute, Moscow State University, 119992, 
Universitetskiy Prospekt, 13, Moscow, Russia
\\
$^{3}$ DAMTP, Centre for Mathematical Sciences, University of
Cambridge, Wilberforce Road, Cambridge CB3 0WA, United Kingdom 
}

\begin{document}

\date{Accepted. Received; in original form}

\pagerange{\pageref{firstpage}--\pageref{lastpage}} \pubyear{2010}

\maketitle

\label{firstpage}

\begin{abstract}

We consider misaligned accretion discs formed after tidal disruption events
occurring  when a star encounters a supermassive rotating black hole.
We use the linear theory of warped accretion discs to 
find the disc shape when the stream produced by the
disrupted star provides a source of mass and angular momentum 
that is misaligned with the black hole. The evolution of
the surface density and   aspect ratio  is  found  from a
one dimensional vertically averaged  model.

We extend previous work which assumed a  quasi-stationary disc
to allow  unrestricted dynamical propagation of disc tilt and twist 
through time dependent backgrounds. We  consider a  smaller value of 
the viscosity parameter,  $\alpha =0.01,$ finding  the  dynamics varies  significantly.

At early times the  disc  inclination is found to be
nearly uniform at small radii where the aspect ratio is 
large. However, since  torques arise from the Lense-Thirring  effect
and the stream  there is non  uniform precession. We propose 
a simple model for this requiring  only  the  background  surface 
density and aspect ratio.     

At these times  the $\alpha \sim 0.01$  disc  exhibits a new feature. 
An  inclined hot inner  region joins  an outer low inclination
cool region  via  a  thin transition front propagating outwards with
a speed  exceeding  that of bending waves in the cool 
region.  These waves accumulate  where the propagation speeds match producing
an inclination  spike separating   inner and outer discs.
At late times  a sequence of quasi-stationary configurations  approximates  
disc shapes  at  small radii. We discuss observational implications of
our results.

\end{abstract}



\section{Introduction}
The tidal disruption of stars by supermassive black holes (tidal disruption events, TDEs)  is expected to result
in accretion of stellar gas onto them and,
accordingly, in an increase in their activity as suggested by \cite{Hil1975}. 
Over the last two decades or so TDEs have been proposed to explain non-stationary radiation
flares observed in several dozen of galactic centres in different wavebands (from radio to X-rays), typically 
lasting for several years, \citep[see e.g.][]{Kom2015}\footnote{Note that TDEs are also invoked to explain changes between different states of AGNs, see \cite{OK2017}, \cite{OK2018}.}  

\subsection{Background}
Recent understanding of a single TDE is based on the following simple picture proposed in \cite{Lac1982} and \cite{Ree1988}. 
It is assumed that all stars with periastron distances, $R_p$, smaller that so-called tidal disruption radius, $R_{T}$, are
disrupted by the tidal field the  of black hole.
Physically $R_{T}$ is defined by the condition that within a uniform density sphere of radius, $R_{T},$ and mass, 
$M,$ equal to that of the black hole,  the  density is the same as the  mean  density of the unperturbed star.
It is given by equation
(\ref{eq1}) below. When the star is located at  $R_{T},$ tidal forces and self-gravitational forces
are expected to balance on rhe stellar surface with tidal forces being larger when the star is at smaller radii. 
Accordingly a  star moving on a nearly parabolic orbit with  $R_{p} < R_{T}$
is expected to become  gravitationally unbound near periastron, \citep[see e.g.][]{Car1983, 
Car1985, Kho1993b,Iva2001,Iva2003, Lod2009, Gui2013, Gui2015} for different numerical models of this process. 

Since the stellar centre of mass has zero binding energy on a parabolic orbit, stellar material that becomes closer 
 to the black hole at periastron has positive binding energy and will accordingly be gravitationally bound
to it. On the other hane, material which is further from the black hole at periastron  will not be bound to it 
and thus expelled from the black hole. 
Accordingly, approximately one half of the  stellar material will begin to  move around the  black hole 
on highly elliptical orbits with periods determined by the distribution of its mass with binding energy,
which is often assumed to be uniform, \citep[e.g][]{Ree1988}.
Since different stellar fluid  elements return to periastron at different times, there will be a stream of gas
coming to periastron with associated  mass flux, $\dot M_{S}$ equal to
${(m/ 3P_{min})}({t/P_{min}})^{-5/3}$, where $m$ is mass of the star
and $P_{min}$ is the minimal return time to periastron  for
 a gas element, which is nearest to the  black hole at the time of periastron passage, 
which is given by equation (\ref{eq3}) below.
 
Since
the gas stream tends to intersect itself near $R_{p}$, on account of differential Einstein precession,
gas circularises 
there and eventually forms an accretion disc or torus. 
Recent numerical simulations suggest that it takes a time 
$1-10P_{min}$ for an accretion disc structure to be formed near $R_{P}$, \citep[see e.g.][]{Hay2013,Bon2016}.
Initially such a disc,  being in an advective state,
accretes at a superEddington rate. 
 But, after some time (typically, order of a year) has passed,  it evolves into  a 'standard' optically thick, geometrically thin, radiative
accretion disc, \citep[see e.g.][]{She2014} and references therein. 
Such a disc differs, however, from the 
standard stationary accretion disc described in \cite{Sha1973} by having a free outer boundary,
its non-stationary nature and mass and energy input being provided  by the stream.  

An important aspect of the problem arises when the black hole rotates. 
Apart from modification of  
the disc spectrum,
 issues related to possible jet formation through the Blandford $\&$Znajek process and modification
of the process of tidal disruption itself \citep[e.g.][and references therein]{Iva2006}, 
 black hole rotation 
could induce a  non-trivial geometrical  structure of the disc.
Indeed, there is no any reason for suggesting that the 
black hole equatorial plane should coincide with orbital plane of the star.
In general, they should be inclined
with respect to each other by angle order of unity. Therefore, at least initially, the disc can be inclined
with respect to the equatorial plane and undergo precession due to the action of the Lense-Thirring torque,
\citep[see e.g.][]{Sto2012,Fra2016}.

In previous  work on this problem \cite{Xiang2016} pointed out that, rather than assuming a free precession,  the geometrical form of the disc 
should be determined, at least, at late times incorporating the dynamical action of the stream, which transfers 
components of angular momentum parallel to the equatorial plane of the black hole, and, therefore, 
tends to 'push' the disc out of this plane.
 Although the mass flux in the stream, and, accordingly,
the flux of angular momentum sharply decay with time, so does mass of the disc due to accretion 
onto the black hole.
On account of this it turns out that the  action of the stream may be important for long times, at least order of
a hundred of $P_{min}.$

In \cite{Xiang2016} the geometrical form of the disc
was estimated using two complementary approaches. 
The first one was through performing SPH simulations of a disc
inclined with respected to the equatorial plane that was impacted  by a source stream of incoming gas 
with appropriate specific angular momentum.
 The second approach  was based on the linear theory of twisted tilted accretion discs 
 \citep[e.g.][]{Pap1983, Pap1995}.
  A modified   form of  an equation derived in \cite{Iva1997} for stationary
configurations of twisted tilted  discs in the  gravitational field of a rotating black hole 
which  incorporates  a source
term due to the presence of the stream was  solved numerically.

 Note that although this equation  does not explicitly contain time derivatives of variables describing  the disc' tilt and twist, time
dependence is present in the solutions arising from  the time dependent source term and the variables characterising
the background  state of the disc fed by the stream.  
This background state refers to the disc obtained when the orbital plane of the stream and the
equatorial plane of the black hole are aligned.
The background state is characterised  by the 'opening angle' $\delta$ defined as the ratio
of disc  semi-thickness $H$ to  the local radius, $R$, and the  surface density $\Sigma.$
The background state variables 
were determined  using a modification of the publicly available code NIRVANA to  take into account mass and
energy input  due to the stream.
 One dimensional time dependent calculations  to find the evolution of such discs were undertaken
for two characteristic values of the   \citet{Sha1973}  viscosity parameter  $\alpha=0.3$ and $\alpha=0.1$

In \cite{Xiang2016} it was found that disc's inclination could be substantial, especially at the time
of transition from an advective to radiative state $t_{tr}\sim 50-100P_{min}$, when the model of the disc employing 
an $\alpha$ viscosity  is thermally unstable \citep[e.g.][] {Sha1976, Abr1988}. 
 During this transition the disc inclination angle to the black hole equatorial plane, $\beta,$  at the stream impact radius can be as large as $(0.1-0.3)\beta_*$,
  where $\beta_*$ is the inclination of the orbital plane of the stream,
  for the maximal black hole rotation. When black hole rotation is smaller the inclination angle becomes even larger.

The quasi-stationary approach of \cite{Xiang2016} 
in which the background quantities are held constant in time 
while the inclination and twist are allowed to attain a steady state.
does not allow us to consider effects associated with the dynamics
of a twisted disc for  which this approximation fails, which is likely to be the case   at early times, when  characteristic tilt and twist 
propagation and relaxation times are larger than the time elapsed from the beginning of the tidal disruption.
 
 \subsection{ Fully time dependent calculations, inclination evolution and precession for a low viscosity disc}                 
 It is the purpose of this Paper to consider the fully time dependent problem, using a set of dynamical 
equations describing propagation and relaxation of disc's tilt and twist derived elsewhere \citep[see][] {Zhu2011,
Mor2014, Zhu2014}  with the modification that an isothermal
density  distribution  in the  vertical direction is incorporated  instead of the polytropic models
 used in these papers. 
 Note that this set of equations
has been derived in a fully relativistic approach, but using a formal assumption that the absolute value 
of black hole rotational parameter, $a$, is small. Nonetheless, they can be also used to consider cases with
$|a|\sim 1$  when scales much larger than the size of marginally stable orbit are considered as in this Paper.
Also, contrary to  \cite{Xiang2016} we consider a background state with  small $\alpha=0.01$ together with  the larger 
$\alpha=0.1$  and perform calculations of background quantities with an increased resolution. We consider a set 
of different initial computation times and values of $a$. For the initial disc model we mainly consider
a flat disc initially aligned with the equatorial plane at some time $t_{in}$ chosen to be equal to $10P_{min}$ for 
most  of  the  computations.  We  also perform several runs for which the disc is initially aligned with 
the orbital  plane containing the initial stream.

We find that disc's evolution depends significantly on the time  elapsed from the beginning of the process. When
$t/P_{min} ~\le 30$  inclination angles are found to be  nearly uniform over a large range of radii as
was suggested in e.g. \cite{Sto2012}. However, for the adopted values of the  black hole mass and 
 the stream disc  impact radius, the
precession of  a disc  annulus  in this phase is accompanied by a strong evolution of its inclination angle.
 Also the duration of  this stage is less than or is comparable to one precessional period
 at the stream impact location. At moments of time $t/P_{min} ~\ge 50-70$ corresponding 
to onset of the thermal instability the disc dynamics is well described by the quasi-stationary model of \cite{Xiang2016}.

\subsection{Accumulation of bending waves in an outward propagating transition front}
In addition, for case with  $\alpha=0.01$ we have found a qualitatively new effect 
in the form of the formation of a spike in the  inclination
angle distribution at the outer edge of the region inside which the inclination angle is nearly uniform. 
The origin of this spike is related
to the  behaviour of the  disc  semi-thickness at early  evolution times, which is very small far away
from black hole, and relatively large at short distances. 
The interface between the 'cold' and 'hot' phases corresponding
respectively, to relatively small and large values of  the half-thickness moves outwards with a speed, which exceeds the propagation
speed of bending waves at and beyond some radius. At this radius bending waves are accumulated leading to the  formation  of a spike and
eventual numerical instability. 
This instability is regularised by adding some artificial dissipation term acting only in
the vicinity of this radius. This is found  not to influence our solutions in other regions of the computational 
domain.
 Physically, this singular behaviour of our system might also  be regularised by  allowing for non-linear terms in the  
equations governing  inclination, or adding terms of higher order in the expansion in powers of  $\delta.$ 
Although this is technically complex it  is not  expected to affect the global disc evolution and so is beyond the scope of this paper. 

The Paper is organised as follows.  In  Section \ref{Basic}  we introduce basic notations and definitions.  Section \ref{Dyn}
is devoted to our dynamical equations describing  disc tilt and twist (referred hereafter as  the twist equations) in the context of the background models.
The background models are described in section \ref{Background} and the twist equations in section \ref{gov}.  In Section \ref{propfront}
we develop  a simple dynamical model of disc behaviour during the stage when  the inclination angle is nearly uniform and
discuss the  behaviour of the  inclination in the spike region located in the
outward propagating  transition front between an inner hot region and an outer cool region, 
 providing a simple model for it in Section \ref{ODE}.
Results of numerical simulations  of the twist equations are discussed in Section {\ref{numres}.
Finally, we summarise  our results and conclude in Section \ref{Discuss}.

\section{Basic Definitions and Notation}\label{Basic}
Although we use the same problem setup  as in Paper 1,
in this Section we briefly review main definitions and parameters used below
in order to make this Paper
self-contained. For an extensive 
discussion see Paper 1.

We introduce a Cartesian coordinate system $(XYZ)$ with origin at the  black hole
location. The  $(XY)$ plane  coincides  with the equatorial plane of the black hole. 
 The angle between
the $X$ axis and the line of intersection of the plane  containing  the stream  with  the equatorial plane of the black hole is  $\gamma_{*}$, its
relative  inclination angle is $\beta_{*}$.
 The inclination  of the disc mid plane  to the $(X,Y)$ plane 
 at  radius, $R,$ is $\beta(t,R)$ and the angle between the  line of intersection of this  plane with the  $(XY)$ plane  and the $X$ axis is
$\gamma (t, R)$.  Following our previous work we introduce the  complex variables
${\cx W}(t,R)=\beta(t,R)e^{{\rm i}\gamma (t, R)}$ and ${\cx W}_{*}=\beta_{*}e^{{\rm i}\gamma_{*}}$, using
calligraphic letters for complex quantities hereafter.

\subsection{Basic spatial and temporal units associated with the stellar orbit and gas stream}\label{Char}

In what follows for  unit of  distance we use the distance from the black hole to the location  where the stream impacts the disc, $R_S$.
In general, we have $R_S > R_p$, where $R_p$ is periastron distance of the initial stellar orbit. \footnote{ Note that
when the disc
is significantly inclined with respect to the plane of the  stellar orbit we have typically $R_S \sim R_p$.}.
We express $R_S$ and $R_P$  in units of  the tidal radius   
\begin{equation}
R_T=\left({{M/ m}}\right)^{1/3} R_*= 7 \cdot 10^{12}M_{6}^{1/3}cm=46M_6^{-2/3}R_g,
\label{eq1}
\end{equation}
through  $R_S=R_{T}/B_S$ and $R_p=R_{T}/B_p.$
Here $M$ and $m$ are the masses of black hole and star, respectively,
$R_*$ is the stellar radius,
$M_6=M/10^6M_{\odot}$, and we define the  gravitational radius as $R_{g}=GM/c^2$,
where $c$  and $G$ are the speed of light and the gravitational constant.

\noindent We  shall assume below that $m$ and $R_*$ have  Solar values,  $M_6=1$ and $B_p=7/4.5\approx 1.55$ 
and $B_S=0.5B_p=7/9$.

An important  characteristic time  corresponds to the minimum return time of gas in the stream to periastron after the disruption of the star. 
This is given by
\begin{equation}
P_{min}={\pi \over \sqrt{2}}(R_p/ R_*)^{3}\left(m/M\right)^{1/2} t_*= 3.5\cdot 10^6M_6^{1/2}B_p^{-3}s\approx 9.44\cdot 10^5s,
\label{eq3}
\end{equation}
where  $t_*= R^{3/2}_*/(Gm)^{1/2}$ and  the last equality expresses $P_{min}$  in terms of   dimensionless parameters of interest. 

For times after the tidal disruption  that exceed   $P_{min},$  the disc gains matter from the stream at a rate
\begin{equation}
\dot M_S={m\over 3 P_{min}}\left({t\over P_{min}}\right )^{-5/3}={1.9\cdot 10^{26}}B_p^{3}\left({t\over P_{min}}\right)^{-5/3}g/s
\approx 10^{26}B_p^{3}\left({t\over P_{min}}\right)^{-5/3}g/s,
\label{eq4}
\end{equation}
\footnote{ 
Note that at sufficiently early times $\sim 10-20P_{min},$ $\dot M_{S}$  could have a more complicated time dependence, with its logarithmic derivative with respect to time ranging from $-1$ to $-2$, see Fig. 1 of \cite{Wu2018}. At those times it is better to use numerically determined  values of $\dot M_{S}$ rather than the form given by (\ref{eq4}). However, we believe that this wouldn't 
alter the  qualitative form of the  results of our paper at later times, as was found
when the form of the energy input due to the stream impact was significantly changed
as mentioned in Section 3.1. These aspects should be investigated further in future studies. }

\noindent As has  been discussed by a number of authors \citep[see e.g.][]{Hay2013, Bon2016}   an accretion disc can be formed from the material returning 
to periastron after  a time order of a few $P_{min}$.
A precise duration for  this, so called circularisation stage is
difficult to calculate, therefore, we start numerical calculations   of  the evolution of  the disc  tilt and twist at  a time
$t_{in}=10P_{min}$ after tidal disruption for  the main part of our numerical work,.
We assume  that the disc at this time is very thin  and  either lies at the equatorial plane with ${\cx W}_{*}=0$ or it coincides
with the plane containing the stream with   ${\cx W}_{*}=\beta_*$. 
In order to investigate the dependence of our results on $t_{in},$ we  performed  several  calculations
with $t_{in}=5P_{min}$ and $15P_{min}$.

We showed in Paper 1 that in the linear approximation the dynamical action of the stream on the disc can be approximately described as 
providing a source of angular momentum with components   corresponding to those of the stream, thus
being perpendicular to the plane inclined at an  angle $\beta_*$ to the the equatorial plane of the 
black hole. The corresponding complex torque acting on the disc has the form  
\begin{equation}
 {\cx T}= \sqrt{GMR_S}\dot M_S ({\cx W}_{*} - {\cx W}),
\label{eq4n2}
\end{equation}
where $ {\cx T}=i\dot L_X -\dot L_Y ,$ with  
$\dot L_X$  and $\dot L_Y $
being the components of the torque  in the $X$ and $Y$ directions.
As in Paper 1 we assume that the action of this torque is concentrated 
in a small region of size $\Delta $ around $R_{S}$ with radial dependence being proportional to a
Gaussian such that  ${\cx T}\propto \delta_{\Delta} = (R_s/\Delta)(2\pi)^{-1/2}e^{-(R-R_{s})^2/(2\Delta^{2})}$.
 In our numerical and analytic work we use $\Delta /R_S=0.01$ and check that other values of $\Delta < R_{S}$ leave our results practically unchanged. 
Using these assumptions and definitions we write
\begin{equation}
 {\cx T}= 2\pi \int dR \Sigma R\sqrt{GMR} \dot {\cx W}_{S}   \label{eq4na}
\end{equation}
where the disc  surface
density is  $\Sigma,$ 
the integral is taken over the disc and
$\dot {\cx W}_{S}$ has the form
\begin{equation}
\dot {\cx W}_{S}={\dot M_S\over 2\pi \Sigma(R_S) R_S^2}({\cx W}_{*} - {\cx W})\delta_{\Delta},
\label{eq11}
\end{equation}
where ${\cx W}$ is evaluated at $R=R_S$.

Apart from the torque provided by the stream there is a torque arising from the rotating black hole 
due to the frame dragging effect. Sufficiently far from the event horizon it can be approximately realized
through the action of the gravitomagnetic force\footnote{For a derivation of  the gravitomagnetic force see 
e.g. \cite{Rug2002} or \cite{Mash2003}. As discussed   there, for this approach to be  generally valid non-linear terms in the Einstein 
equations are neglected,  gravitational forces should be stationary and  gas velocities should be smaller
than the speed of light.
However, we would like to stress that equation (\ref{eq9}) below provides a fully relativistic expression 
for Lense-Thirring precession when the  black hole rotation parameter $a$ is small and accordingly only terms linear in $a$ are
taken into account \citep[see e.g.][] {Zhu2011}.  Additional constraints as indicated above  then do not apply.}
This results in the  precession of a  ring  that is inclined 
with respect to the  equatorial plane of the black hole around the rotation axis with  the  Lense-Thirring
frequency
\begin{equation}
\Omega_{LT}=2a{G^2M^2\over c^3 R^3},
\label{eq9}
\end{equation}
where $a$ is black hole rotation parameter. This  parameter lies in  the range $-1 \ge a \ge 1$, with negative values
corresponding to the situation when black hole rotates in the direction opposite to that of the disc orbital motion and $a=0$
corresponding to a non-rotating black hole.

\section{Dynamical equations}\label{Dyn}

 We study the  evolution of ${\cx W}$ in the linear approximation in which the
angle $\beta$ is assumed to be small in magnitude.	Thus, we solve numerically a set of  linear 
equations describing the joint dynamical  evolution of  ${\cx W}$ and  an additional variable ${\cx B}$ 
(see equations (\ref{eq_B}), (\ref{eqW}) below) with coefficients, which are 
themselves are functions of time. Their time dependence is determined by the evolution of 
disc aspect ratio $\delta=H/R$,  with $H$ being the local semi-thickness and the surface density $\Sigma$ described
hereafter as background quantities.

\subsection{Model of  the evolution of the background quantities}\label{Background}

In order to determine the time dependence of $\delta$  and $\Sigma$ we adopt  a one dimensional model 
for which the disc state variables  depend only on $R$  and time $,t.$   We  take
 account of  the influence of the stream  by assuming that there is  an input of mass and energy to the disc at
$R=R_S$  at the rates $\dot M_S$ and  $0.5f_{ st}(GM/ R_S)\dot M_S$, respectively.
Here $f_{st}=1$ corresponds to the maximum available rate of kinetic energy dissipation. 
For most calculations we adopted $f_{st}=1/2.$ Checks made with this reduced by an order of magnitude
indicate little qualitative change to the results. 
 The gravitational potential is taken to have  the Paczynski-Wiita form, see \cite{Pac1980}: $\Phi=
-{GM/ (R-2GM/c^2)}$ and we neglect the  self-gravity of the disc. The  state variables  of interest are either taken in  the disc
mid-plane or are integrated over the  vertical direction and averaged over azimuthal angle. Their dynamical evolution 
is described by equations (29)~ -~ (38) of \cite{Xiang2016}.

We integrate these equations numerically for sufficiently long times (typically, $\sim 100-20P_{min}$) to reveal 
the transition in the form of the  evolution of the disc from the  initial advection dominated slim disc regime with $\delta \sim 1$
to later stages of evolution, when the accretion rate becomes smaller than the Eddington limit and energy transport by advection
becomes negligible with $\delta $ is expected to become quite small, $\delta \sim 10^{-3}$. 

We adopt the usual
assumption that the evolution of the disc is  governed by a turbulent viscosity modelled through
the $\alpha$ prescription, where kinematic viscosity $\nu $ takes the form
\begin{equation}
\nu =\alpha P/(R \rho |d\Omega/dR|),
\label{eq5}
\end{equation}
where the viscosity parameter $\alpha < 1$ is a constant, the  pressure $P$ and density $\rho$ are evaluated in the disc
mid-plane and $\Omega=\sqrt{GM/( R (R-2GM/c^2)^2)}$ is angular frequency of a particle on circular orbit
under the Paczynski-Wiita  potential\footnote{We remark  that SPH simulations of twisted discs employing an
$\alpha$ prescription give results which are in agreement with MHD  simulations 
\citep[see][]{Nea2016}. Also,
it is of  interest  to point out that \cite{Nix2015} has shown,  for  a particular twisted disc model,  that an
$\alpha $ prescription seems  adequate to describe the  luminosity changes of Her X-1 occurring with the well known $35$
day period.}. In order to ensure that the calculated value of  $\delta$ did not exceed  unity, viscous  mid-plane heating was quenched 
under the ad hoc assumption that energy would drive an outflow under extreme circumstances, \citep[see][]{Xiang2016}.
The equation of state is assumed to be that of  a mixture of ideal gas and radiation, the radiative fluxes in  the radial and vertical direction are given by equations (32) and (37) of \cite{Xiang2016}.
As discussed in 
\cite{Xiang2016},  the  disc semi-thickness, $H,$ can be found with adequate accuracy from
\begin{equation}
H =\sqrt{2P/(\rho \Omega^2)},
\label{eq5n}
\end{equation} 

In general, we solve equations determining $\Sigma (t, R)$ and
$\delta (t, R)$ numerically, using a one-dimensional version of the publicly available hydrodynamical code 
NIRVANA,  see  Section 5 of \\\
\cite{Xiang2016} for details, and use the 
values obtained as input for our analytic model for  the evolution of the disc  tilt and twist.
We choose the  unit  of  surface density, $\Sigma_0$, to be determined by the stellar mass and the stream
impact distance according to $\Sigma_0=m/( 2\pi R_S^2).$
We then define  the  dimensionless surface density $\tilde \Sigma =\Sigma /\Sigma_0.$

\begin{figure}
\vspace{0.5cm}
\begin{center}
\vspace{10cm}\hspace{-10cm}\includegraphics{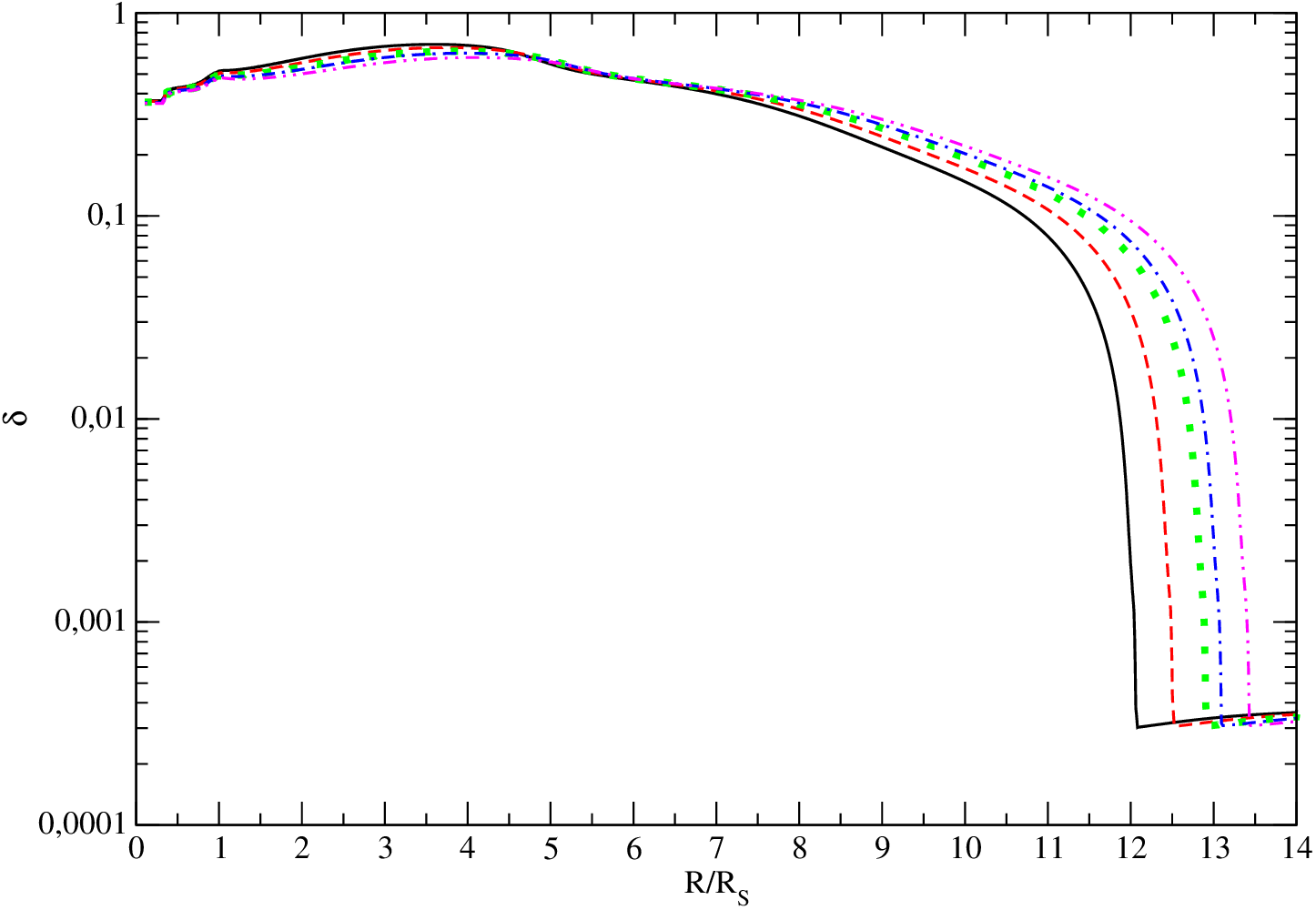}
\end{center}
\vspace{-0.5cm}
\caption{The dependence of  disc  aspect ratio $\delta$ on radius calculated
for the  low viscosity run with $\alpha=0.01$. Solid, dashed, dotted, dot-dashed 
and dot-dot-dashed curves correspond to $t/P_{min}\approx 11,12,13,14,15$,
respectively.}
\label{Fig1}
\vspace{-0.0cm}
\end{figure}

\begin{figure}
\begin{center}
\vspace{10cm}\hspace{-10cm}\includegraphics{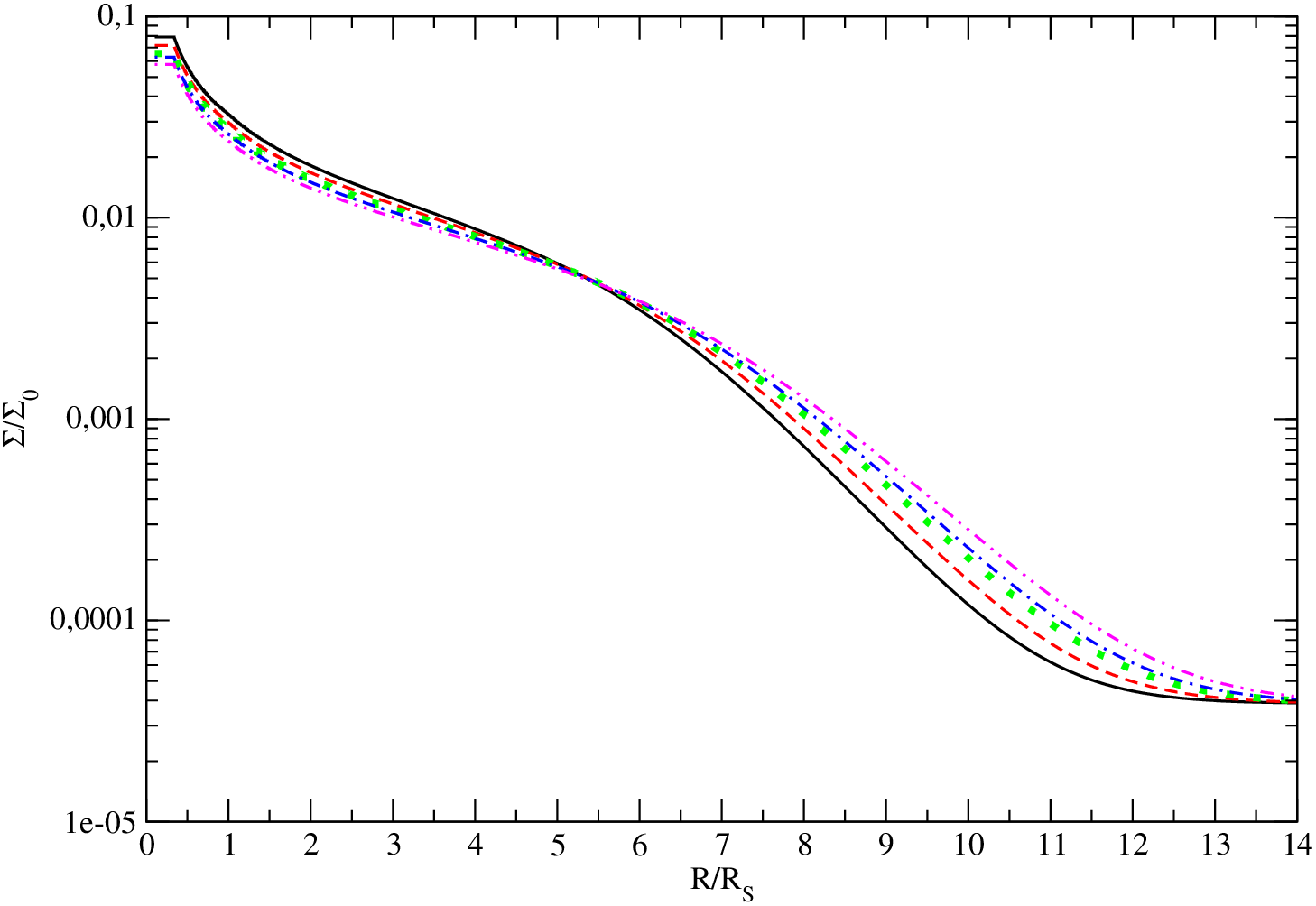}
\end{center}
\vspace{1cm}
\caption{Same as Fig. \ref{Fig1}, but for the disc's surface density expressed in units of its characteristic value $\Sigma_0={m/( 2\pi R_S^2)}$.}
\label{Fig2}
\vspace{-0.5cm}
\end{figure}

\begin{figure}
\begin{center}
\vspace{10cm}\hspace{-10cm}\includegraphics{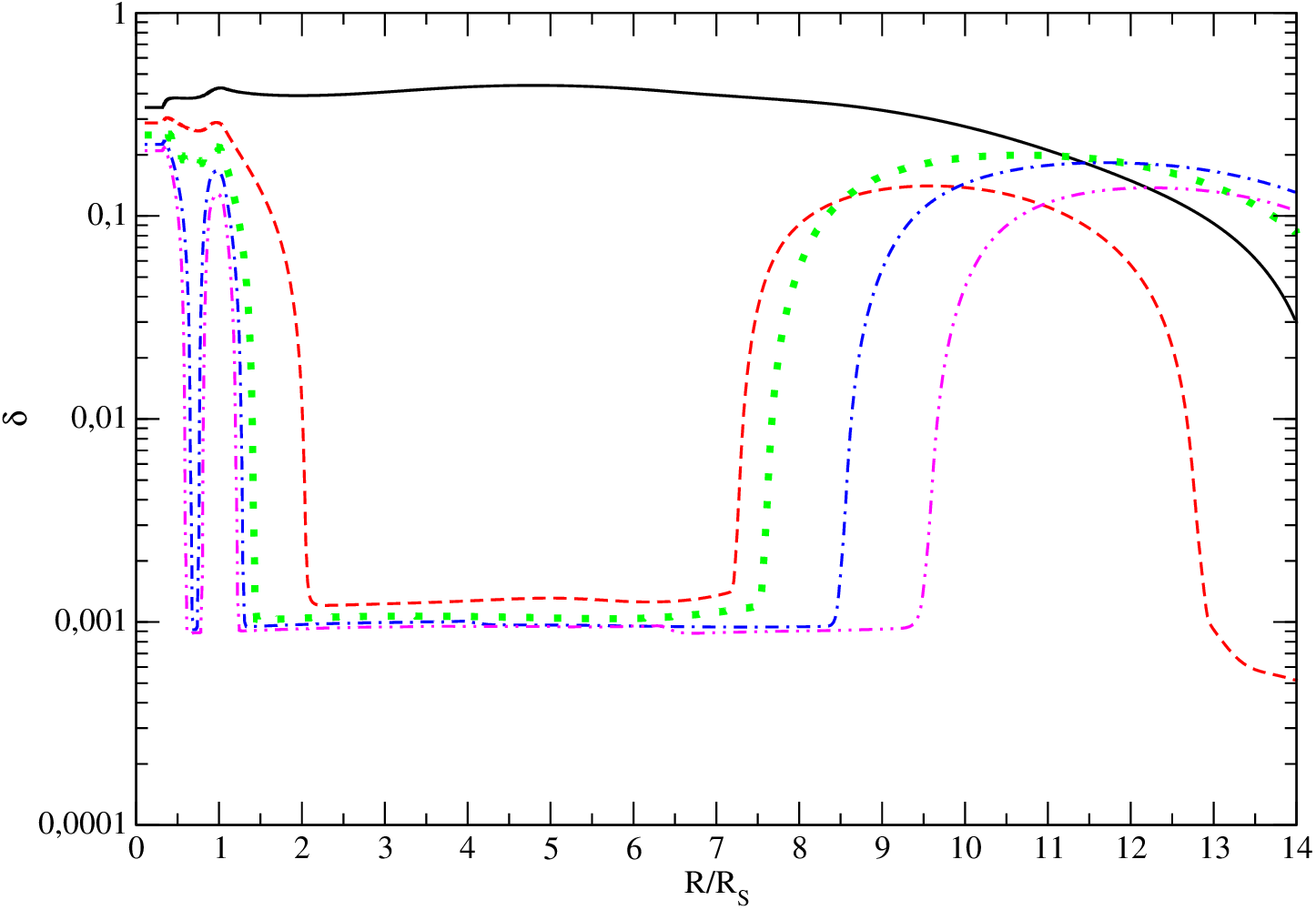}
\end{center}
\vspace{1cm}
\caption{Same as Fig. \ref{Fig1}, but at later times.  Solid, dashed, dotted, dot-dashed and dot-dot-dashed curves correspond to $t/P_{min}\approx 20,40,60,80,100$,
respectively. }
\label{Fig3}
\vspace{-0.5cm}
\end{figure}

\vspace{1cm}
\begin{figure}
\vspace{0cm}
\begin{center}
\vspace{10cm}\hspace{-10cm}\includegraphics{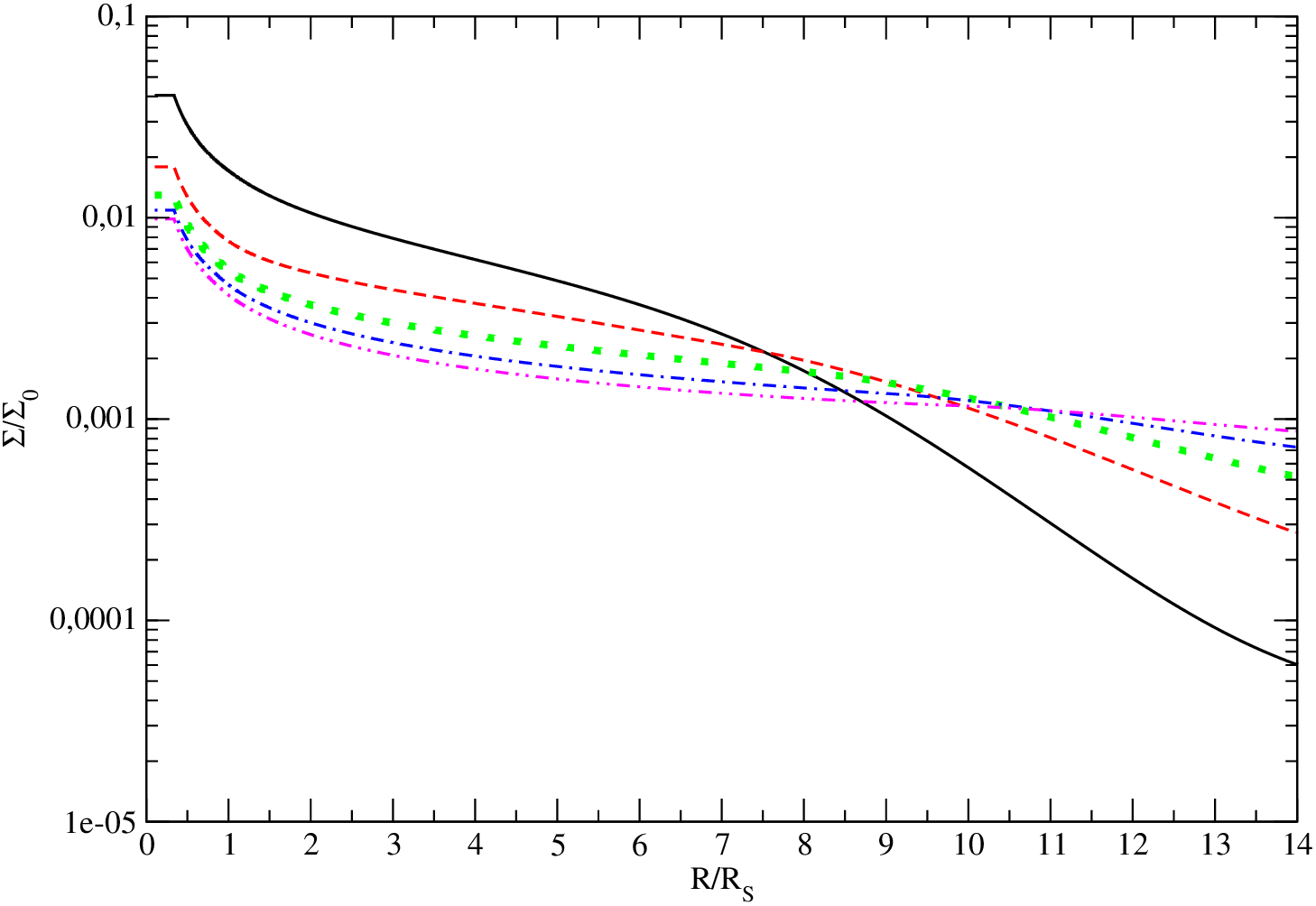}
\end{center}
\vspace{0cm}
\caption{Same as Fig. \ref{Fig3}, but for the surface density  expressed in units of its characteristic value $\Sigma_0={m/( 2\pi R_S^2)}$.}
\label{Fig4}
\vspace{-0.5cm}
\end{figure}

\begin{figure}
\begin{center}
\vspace{1cm}
\vspace{10cm}\hspace{-10cm}\includegraphics{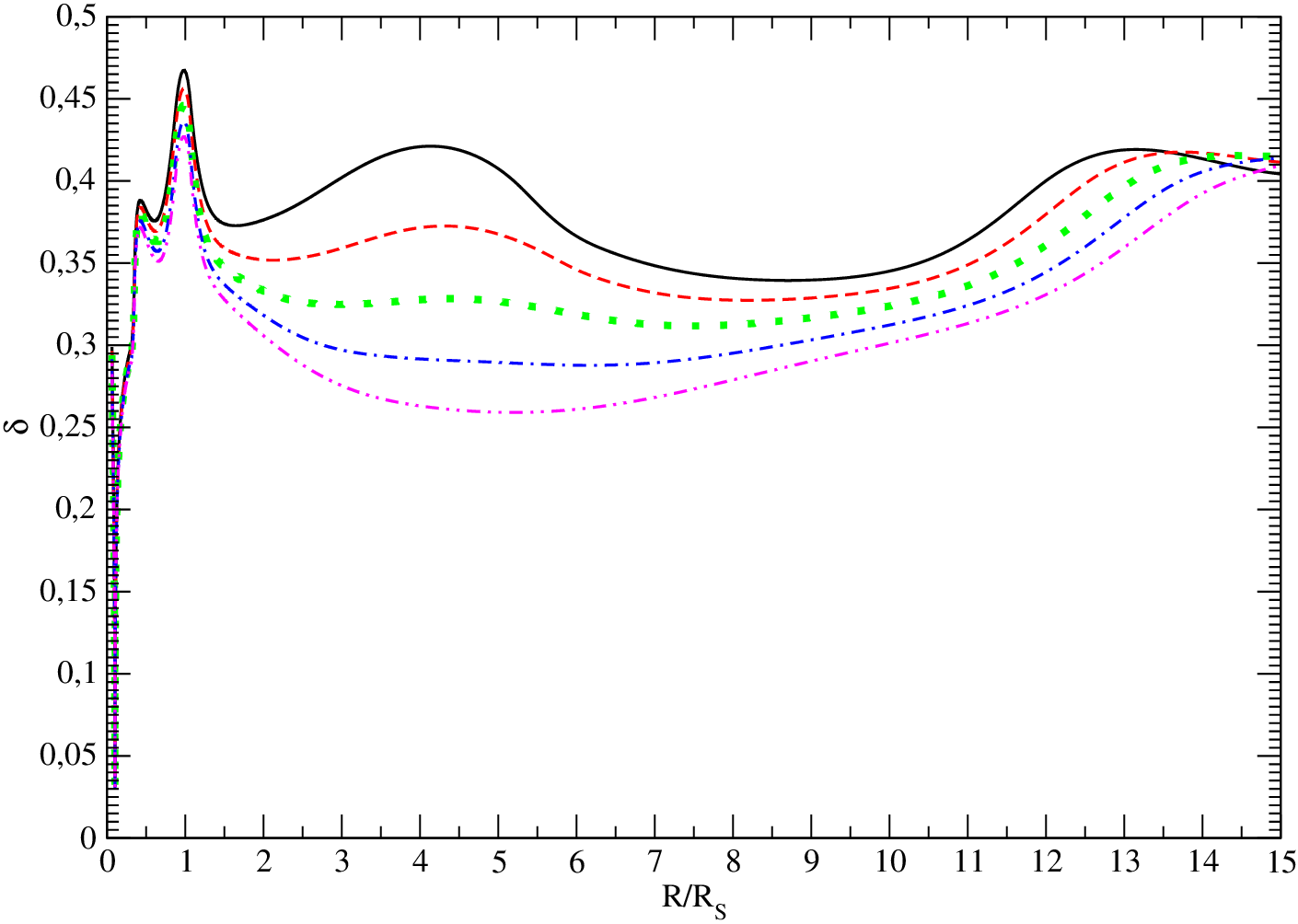}
\end{center}
\vspace{1cm}
\caption{Same as Fig. \ref{Fig1}, but  for the run with larger $\alpha=0.1$.}
\label{Fig5}
\vspace{-0.0cm}
\end{figure}

\begin{figure}
\begin{center}
\vspace{10cm}\hspace{-10cm}\includegraphics{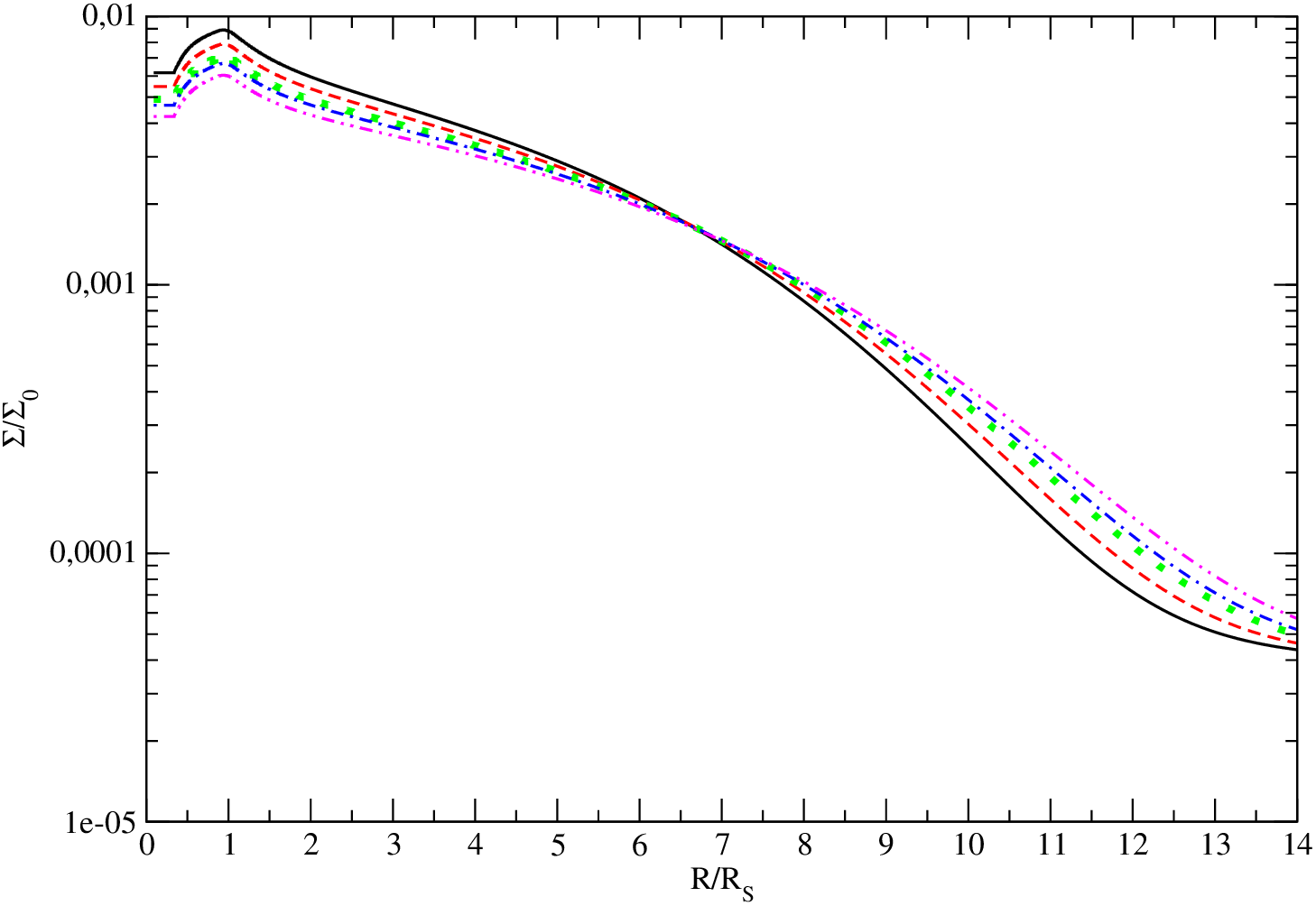}
\end{center}
\vspace{1cm}
\caption{Same as Fig. \ref{Fig5}, but for the surface density expressed in units of its characteristic value $\Sigma_0={m/( 2\pi R_S^2)}$.}
\label{Fig6}
\vspace{-0.5cm}
\end{figure}

\begin{figure}
\begin{center}
\vspace{10cm}\hspace{-10cm}\includegraphics{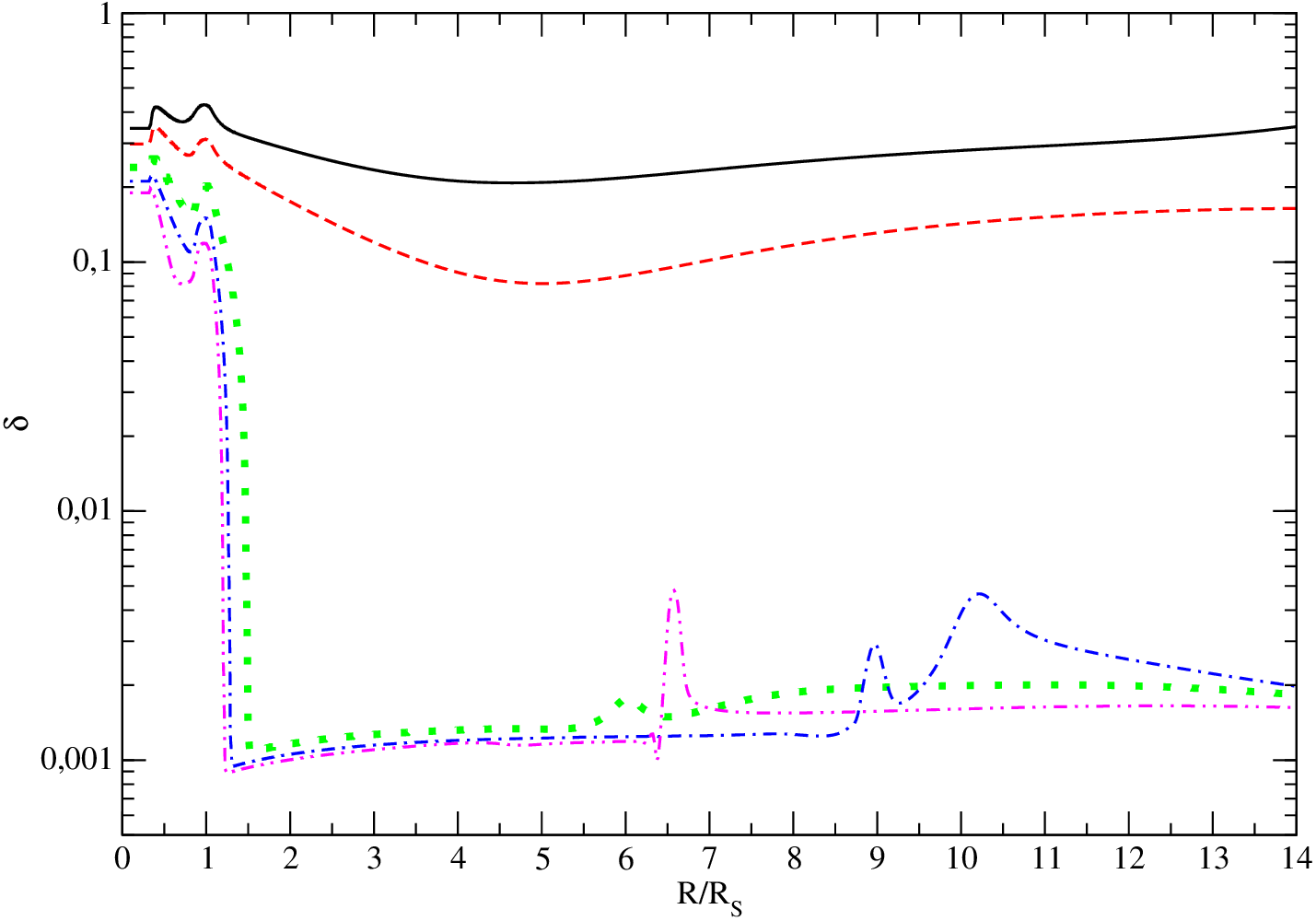}
\end{center}
\vspace{1cm}
\caption{Same as Fig. \ref{Fig3}, but for the run with $\alpha=0.1$.}
\label{Fig7}
\vspace{-0.5cm}
\end{figure}

\begin{figure}
\begin{center}
\vspace{10cm}\hspace{-10cm}\includegraphics{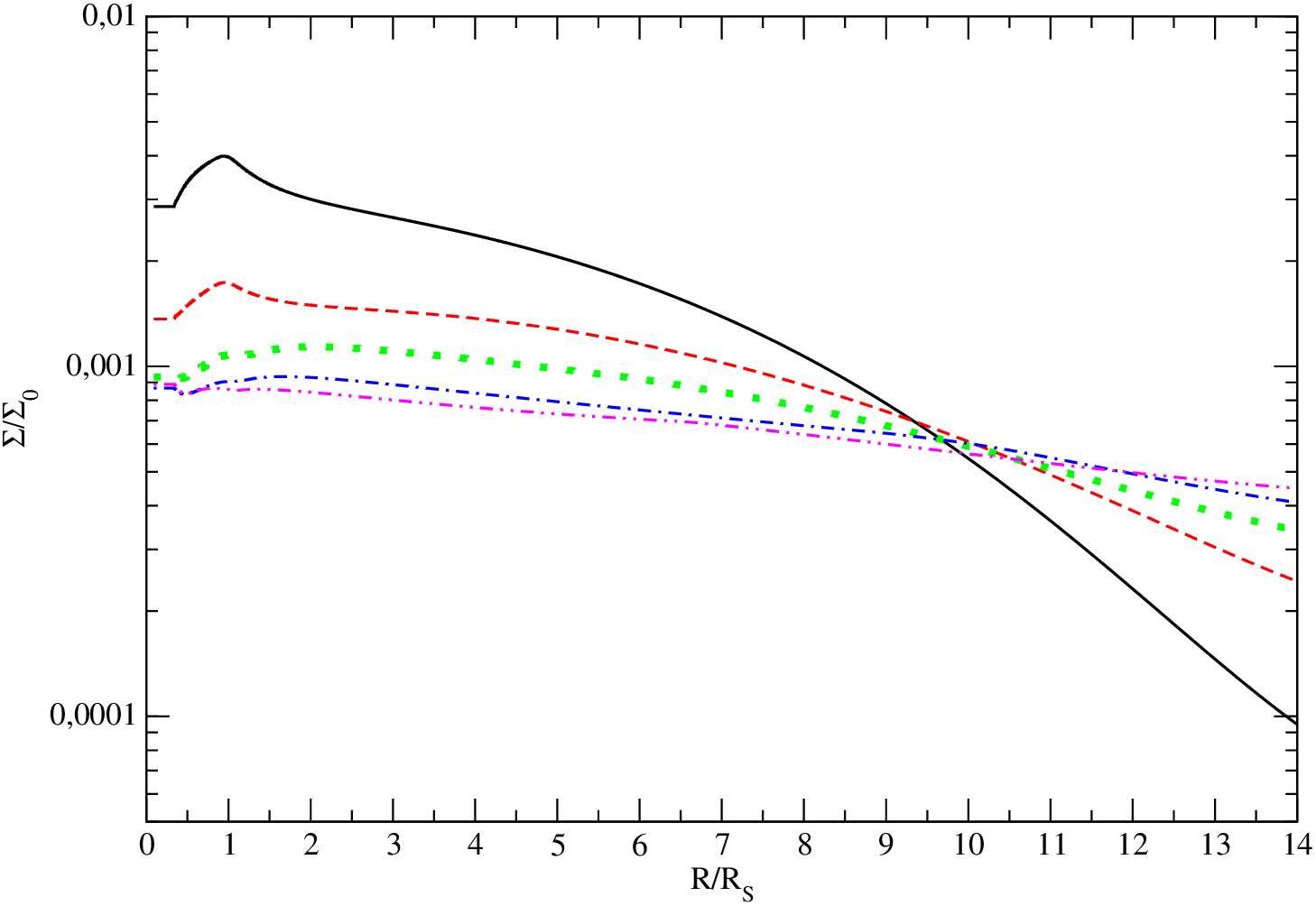}
\end{center}
\vspace{0cm}
\caption{Same as Fig. \ref{Fig7}, but for the  surface density expressed in units of its characteristic value $\Sigma_0={m/( 2\pi R_S^2)}$.}
\label{Fig8}
\vspace{-0.0cm}
\end{figure}

For reference we provide some results of our calculations of the background quantities in Figs. \ref{Fig1}-\ref{Fig8}
for cases $\alpha =0.1,$ and $\alpha =0.01.$
Values of $\alpha =0.1$ and $\alpha =0.3$ were  considered in \cite{Xiang2016}. 
 Figs \ref{Fig1}-\ref{Fig4} illustrate  the case with small $\alpha=0.01,$
while Figs. \ref{Fig5}-\ref{Fig8} correspond to the case with $\alpha=0.1$.

In Figs. \ref{Fig1} and \ref{Fig2} the forms of  the disc aspect
ratio $\delta $ and surface density $\Sigma $ are shown as functions of $R$ at five, relatively
small, values of time, $t/P_{min}\approx 11, 12, 13, 14$ and $15$. We see that while the surface 
density distributions at these  different moments of time are quite similar, and are  
 monotonically decreasing with $R$ and  smooth functions of time, the distributions
of $\delta$ behave in a more complicated way. They all describe a geometrically thick disc 
for  $R/R_S \sim < 10$ joined by a transition region in which $\delta $ sharply decreases with $R.$ 
This joins to a very thin outer disc with $\delta < 10^{-3}$. The boundary between the transition region 
and the thin disc is very sharp having the character of a shock front. It moves outwards with time. 
We find  that the propagation 
speed of this boundary referred hereafter to as the speed of transition front, $v_f$, exceeds the sound 
speed in the outer disc.  The propagation speed may be estimated from the following simple arguments.
Namely, it is well known that a characteristic time scale of the development of thermal instability $t_{Th} \sim \alpha^{-1} \Omega^{-1}$. On the other hand, from Fig. \ref{Fig1} it follows that a typical radial extent of the transition front $\Delta R$ is of the order of a typical disc thickness behind the front $\sim 0.1R$. It is expected that $v_{f}\sim \Delta R/t_{th}=0.1\alpha v_{K}=10^{-2}v_{k}$, where $v_{K}=R\Omega$ is Keplerian velocity. This estimate can exceed the sound speed in the outer cold disc and  is consistent with our numerical results, which give  a value of $v_{f}$ twice as  large. The very outermost part of the transition region may thus take on the character of a shock.

Figs. \ref{Fig3} and \ref{Fig4} show the functional forms  of $\delta $ and $\Sigma $ at later
 times $t/P_{min}=20,40,60,80$ and $100.$ 
One can see in Fig. \ref{Fig3} that at times $> 40P_{min}$ the aspect ratio drops to very small values for $2 < R/R_S < 6-7$. 
A region around $R/R_S=1 $ always has a large value of $\delta $. This is due to heating of disc material by the stream. 
Note that there is also a region of large $\delta $ at larger radii, $R/R_S > 6-9$.
This is because outward  transport of angular momentum and mass has caused the surface density to significantly  increase at the later times
relative to its initial value. This  results in these regions increasing their optical thickness, heating up and undergoing thermal instability,

\noindent  As seen from Fig. \ref{Fig4} the functional form  of the  surface density as a function of radius  gets flatter
with time. 

Results for the case of relatively large $\alpha=0.1$ are illustrated in
Figs. \ref{Fig5} and \ref{Fig6}. These  show the same quantities  at the same times as in respectively  Figs \ref{Fig1}
and \ref{Fig2}.  However, the plots in  Fig. \ref{Fig5} show that
contrary to the case of small $\alpha,$ at the early times considered $\delta $ has relatively large
values throughout the computational domain with a slight bump at $R/R_S \sim 1$ due to heating 
by the stream. In fact this simulation does exhibit an outward propagating front similar to that seen in Fig. \ref{Fig1}.
However, it reaches the outer boundary after a time $\sim 2.6$ which is before the earliest time illustrated 
here  \citep[see][]{Xiang2016}.
 Fig. \ref{Fig6} shows that surface density profiles are also close to each other,
  they all have a maximum at at $R/R_S=1$ and smoothly decrease towards larger radii.

Fig. \ref{Fig7} shows that at for times  exceeding on the order of  $60P_{min}$  $\delta $ drops to 
small values at radii slightly larger than $R_S.$  Note that there are some indications of the
propagation of thermal waves of a moderate amplitude at large radii and late
times as can be seen from the curves corresponding to $t=80$ and $100$. Fig. \ref{Fig8} shows
that  as for the case of small $\alpha $ the surface density distribution gets flatter at
late times.        
\begin{figure}
\begin{center}
\vspace{10cm}\hspace{-10cm}\includegraphics{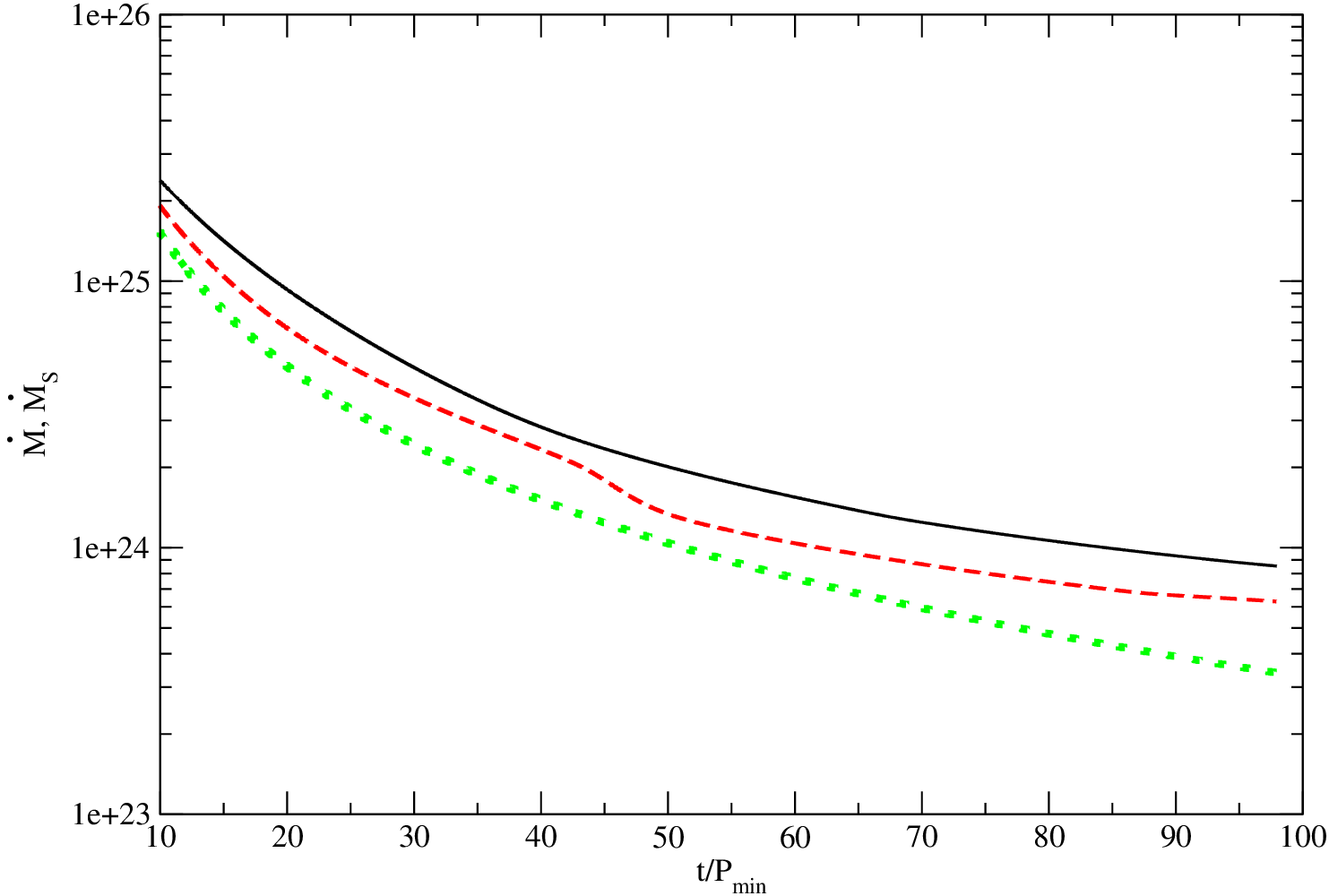}
\end{center}
\vspace{0cm}
\caption{Accretion rates  in g/s as functions of time are shown as solid and dashed curve for 
$\alpha=0.01$ and $\alpha=0.1$, respectively. The dotted curve shows the mass flow in the stream given by equation (\ref{eq4}).}
\label{Fig8n}
\vspace{-0.0cm}
\end{figure}

Finally, in Fig. \ref{Fig8n} we show the dependence of accretion rate, ${\dot M},$ determined as the magnitude
of the mass inflow rate at  the innermost  grid point of our computational 
domain  on time in comparison with the stream mass flow rate $\dot M_S$ given by equation (\ref{eq4}), 
for both  $\alpha=0.01$ and $\alpha=0.1$. As seen from this Figure both accretion rates deviate from $\dot M_S$ at late
times taking on larger values, with the curve corresponding to $\alpha=0.01$ showing a larger deviation.

\subsection{The governing equations for a twisted tilted disc}\label{gov}
The dynamical equations we adopt  for the propagation of disc tilt and twist
are the same as were used by \cite{Mor2014} and \cite{Zhu2014} apart from the fact that we here
 allow for an isothermal  
 density structure in the vertical direction which takes the form
\begin{equation}
 \varrho(R,z,t) =\rho(R,t)\exp{(-z^2/(2H^2))},
\label{en1}
\end{equation}
where the  mid-plane density  $\rho$ and disc semi-thickness $H$ are, in general,
functions of radius $R$ and time. 
The dynamical equations we use can be easily obtained
from expressions contained in \cite{Zhu2011},
as  outlined in appendix \ref{outline}. 
Here we only discuss their form and some basic properties.

As in  previous our work, the $\alpha $ parameter entering our
 dynamical equations is defined through 
the relation
\begin{equation}
\nu=\alpha \delta^2 \sqrt{GMR},
\label{en1a}
\end{equation}
where $\nu$ is the  kinematic viscosity defined above.
But note that this definition differs from that made through (\ref{eq5n})
(see below for a reconciliation).

There are, in general, two independent variables that specify the orientation of the disc, 
the variable ${\cx W}$ introduced above and 
an additional variable ${\cx B}$, which  describes
the deviation of the trajectories of disc particles
from circular form due to the presence of the disc tilt and twist
(see equation (\ref{def_B}).
Note that some authors (e.g. \cite{Dem1997}) use 
${\cx A}=-{i\over 2}{\cx B}$ instead of ${\cx B}$.

The dynamical equation for ${\cx B}$ has the form 
\begin{equation}
\frac{\partial{\cx B}}{\partial \tau} =  \frac{1}{2} \left \lbrace
\left [ 1 + \frac{\kappa^2}{(i-\alpha)^2 \Omega^2} \right ]
(i-\alpha)\Omega {\cx B} - \left [ (i+\alpha)U^\varphi \Omega
- \frac{3i\alpha}{i-\alpha} K_1 (U^\tau)^2 U^\varphi \tilde
\Omega \right ] K_1 \frac{\partial{\cx W}}{\partial R} \right
\rbrace, \label{eq_B}
\end{equation}
where we use geometrical units, expressing spatial and temporal scales in terms of
 $GM/c^2$ and $GM/c^3$, respectively. We define the time  expressed in these units
 as $\tau =  t c^3/(GM).$  
  Expressed in these units,
$\Omega=R^{-3/2}$ is the angular frequency of a
free particle moving on a circular orbit around a non rotating  black hole, $\kappa$ is the
relativistic epicyclic  frequency, and  $\tilde \Omega $ is another characteristic frequency.
Expressions for  $\kappa $ and $\tilde \Omega$
are given by
\begin{equation}
\kappa = \sqrt{R^{-3} \left (1-\frac{6}{R} \right )}, \quad
\tilde\Omega = \frac{R - 3}{R^2
(R-2)^{1/2}}.\label{eq2011a}
\end{equation}
The quantities $U^{\tau}$ and $U^{\varphi}$  represent two components of
the  four velocity of a free particle moving on a circular orbit 
around a Schwarzschild black hole and are given by
\begin{equation}
U^\varphi = (R-3)^{-1/2},\hspace{2mm} {\rm and} \quad
U^\tau = \left ( \frac{R-2}{R-3} \right
)^{1/2}. \label{eq2011b}
\end{equation} 
The metric components  
of a Schwarzschild black hole in the usual coordinate system are defined by
$K_1=\sqrt{R-2)/R}$.

The dynamical equation for ${\cx W}$ has the form
\begin{multline}
\frac{\partial {\cx W}}{\partial \tau} - i\Omega_\mathrm{LT} {\cx W} +
K_1^2 \left \{ \frac{U^r}{U^\tau} + \frac{3}{2}\alpha K_1
U^\tau U^\varphi \delta^2 \right \}
\frac{\partial {\cx W}}{\partial R} = \\
K_1\left (\, 2R^2 U^\tau U^\varphi \Sigma \, \right )^{-1}
\frac{\partial}{\partial R} \left \{ \Sigma R^3 K_1 \delta^2
U^\varphi \left [\,\,\left ( i+\alpha \right ) {\cx B} +
\alpha_1 K_1 U^\varphi \frac{\partial{\cx
W}}{\partial R}\,\right ] \right \}+\dot {\cx W}_S, \label{eqW}
\end{multline}
where $\dot {\cx W}_{S}$ is given by equation (\ref{eq11}), in our dimensionless units the Lense-Thirring frequency takes the form 
\begin{equation}
\label{Om_LT}
\Omega_{LT} = \frac{2a}{R^3},
\end{equation}
and $U^{r}$ is a component of the four-velocity describing the  drift of gas elements
 in the  radial direction due to viscosity.
It is given by the expression
\begin{equation} U^{r}=
-3\alpha{(1-{3/ R})\over \sqrt{(1-2/ R)}(1-6/ R)}R^{-1/2}
\Sigma^{-1}
{\partial \over \partial R}\left \lbrace 
{\left({(1-{2/ R})\over (1-{3/ R})}\right)}^{3/2}{\left({H\over R}\right)}^{2}R\Sigma\right\rbrace. 
\label{ur}
\end{equation}

Note that we use a different value of the viscosity parameter, $\alpha_1$, 
in the last term in the brackets
on the right hand side  of equation (\ref{eqW}).
As  will be discussed below we  take $\alpha_1$ to be   larger 
than the formally expected value, $\alpha,$ in the vicinity of certain values of $R$ 
to   deal with a specific  numerical  instability, which is present
in the system when $\alpha $ is sufficiently small.  

Equations (\ref{eq_B}) and (\ref{eqW}) formally describe the dynamical evolution  of  the disc tilt and twist in a fully relativistic 
setting under the assumption that black hole rotation is small, and, accordingly, $|a| \ll 1$. However, 
these equations can also be used in a situation when $|a| \sim 1$  and scales  much larger than the gravitational 
radius are considered as will be done below. 
In this context we note that relativistic effects are more significant for the description of the disc tilt and twist
than for the background state on account of the importance of small deviations from non relativistic Keplerian motion
being able to play an important role in the former case. 
Accordingly the background quantities $\Sigma $ and $\delta$ entering (\ref{eq_B}) and (\ref{eqW}) are calculated 
using  a non-relativistic numerical code and on a limited computational domain. 
Also the definitions of 
$H$ and as noted above,  $\alpha, $ in the equations describing background quantities differ slightly. 
It is easy to see that in order to match definitions of $H$ and $\alpha$ through equations (\ref{eq5n}) and
(\ref{eq5}) and, respectively, through equations (\ref{en1}) and (\ref{en1a}) we should assume that
the values of $H$ and $\alpha $ to be adopted in equations  (\ref{eq_B}) and (\ref{eqW}) are
 slightly less than those  used in the equations for the background quantities,  through the mappings
$H \rightarrow H/\sqrt{2}$ and $\alpha \rightarrow {2\over 3}\alpha$ as will be understood below. 

After these adjustments are made we extrapolate numerically obtained values of $\delta $ and $\Sigma$
 to smaller radii through a procedure described in Appendix \ref{extension}.
 
\noindent  In addition we assume that near the marginally stable 
orbit,  the disc is quasi-stationary and is close to the \cite{Nov1973} solution. As discussed by 
 \cite{Nov1973} both $\delta $ and $\Sigma $ should be proportional to the function
\begin{equation}
D=1-{\sqrt 6\over y}-{\sqrt 3\over 2y}\ln{{(y-\sqrt 3)(3+2\sqrt
2)\over (y+\sqrt 3)}}, \label{eq2011c}
\end{equation}
where $y=\sqrt {R}$  which vanishes   at the marginally stable orbit.
Accordingly in addition to the above mentioned extrapolation, we multiply $\delta $ by the factor
 $D^{3/20}$ and $\Sigma $ by the factor $D^{7/10}$, which are appropriate in the situation where 
the disc is gas pressure dominated with opacity determined by Thomson scattering. Note that the assumption
of gas pressure domination fails at sufficiently early  evolution  times of our system. Nonetheless we  continue to use 
these factors even in this case since a more appropriate renormalisation of $\delta $ and $\Sigma $ was found to
not to influence  our results significantly. 

\subsubsection{The governing equations in the large radius and low viscosity limit}
Equation (\ref{eq_B}) and (\ref{eqW}) can be brought into a much simpler form under the assumptions that $R \gg 1$
and $\alpha = O(1/R) \ll 1$. In this limit (\ref{eq_B})  becomes on expanding each term to leading order in $1/R$ 
\begin{equation}
\dot{\cx B} \equiv \frac{\partial {\cx B}}{\partial \tau} =-{i\over 2R^2}{\partial \over \partial R}{\cx W}-\left(\alpha -{3i\over R}\right)\Omega {\cx B}
\label{e1}
\end{equation}
and (\ref{eqW}) gives 
\begin{equation}
\dot{\cx W}\equiv\frac {\partial {\cx W}}{\partial \tau} ={\delta^2\over 2\xi R}{\partial \over \partial R}\left [\xi R^{2}\left (i{\cx B}+\alpha_1R^{-1/2}{\partial \over 
\partial R}{\cx W}\right)\right ]+i\Omega_{LT}{\cx W}+\dot {\cx W}_S,
\label{e2}
\end{equation}
where $\xi =\Sigma \delta^2 R^{1/2}.$ 
The stationary variant of these equations with
$\dot {\cx B}=\dot {\cx W}=0$ was solved in \cite{Xiang2016}. When $\xi=1$ and $\alpha_1=\dot {\cx W}_S=0$ these equations coincide 
with those derived in \cite{Dem1997}. As was  indicated previously  the first term on the right hand side  of (\ref{e1}) 
results from the influence of pressure on the  motion of gas particles on nearly circular trajectories in the disc,
while the terms in the brackets describe the influence of viscosity and the post Newtonian correction leading 
to Einstein precession of free particles moving around a black hole.   

\subsubsection{Local dispersion relation}
It is instructive to obtain a dispersion relation derived from equations  (\ref{e1}-\ref{e2}). Setting 
$\dot {\cx W}_S=0$ and assuming that both dynamical variables $\propto e^{i(\omega \tau + kR)}$ with 
$kR \gg 1$ so that we may adopt a local  perturbation analysis, we find that local perturbations are governed by the 
 dispersion relation 
\begin{equation}
\left(\omega -\Omega_{LT}-i\gamma_1\right )\left (\omega - {3\Omega /R}-i\alpha \Omega \right)={\delta^2 \Omega^2 (kR)^2\over 4}, 
\label{e3}
\end{equation}
where $\gamma_1=(1/2)\alpha_1\delta^2 (kR)^2\Omega$. 

Sufficiently far from  the black hole the contributions of the  Lense-Thirring frequency and the post-Newtonian correction, contributed by the second term
 in the second bracket on the left hand side of
 (\ref{e3}),  can be neglected. 
 In this case, in the low viscosity limit for which  $\alpha < \delta ,$ the  tilt and twist
propagate as waves with no dispersion and propagation speed equal to one half of the sound speed, 
$Re(\omega)\approx \pm (1/2) \delta \Omega kR,$ \citep[see][]{Pap1995} 
and decay rate approximately equal to $(1/2)(\alpha \Omega +\gamma_1).$
From the condition $\gamma_1 < \alpha \Omega$ we find a condition that the term proportional to 
$\alpha_1$ is unimportant to be
\begin{equation}
\alpha_1 < {2\alpha \over \delta^2 (kR)^2}.
\label{e4}
\end{equation}

\section{The dynamics of tilt and twist 
when an outward propagating front separates an inner hot region from
an outer cool region}\label{propfront}
In this Section we discuss the behaviour of the inclination angle $\beta$ as a function of  radius
$R$ when the background disc model has an outward propagating front that separates an inner hot region
of the disc with  moderately large aspect ratio from an outer low aspect ratio cool  region as illustrated  in  Fig. \ref{Fig1}.
In doing this we first discuss a numerical instability  that may be present in the calculation
of $\beta,$ the physics that drives this and how it can be brought under control.

\subsection{A numerical instability arising from the pile up of outward propagating  short wavelength bending waves at the 
transition front with nearly uniform inclination angle at smaller radii}

\begin{figure}
\centering
\vspace{1cm}
\hspace{-1cm}
\includegraphics[width=8cm,height=8cm]{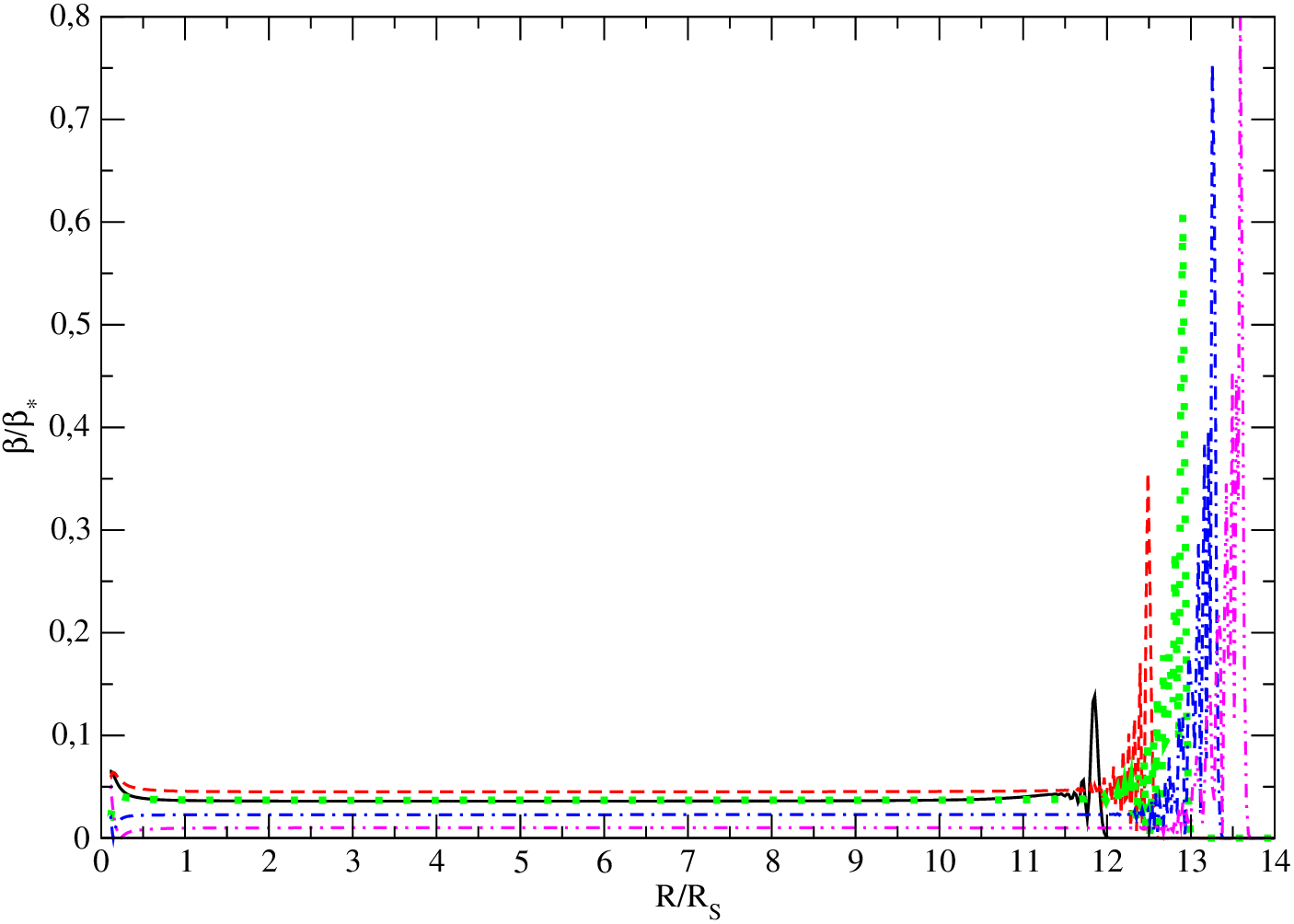}
\hspace{0.5cm}
\includegraphics[width=8cm,height=8cm]{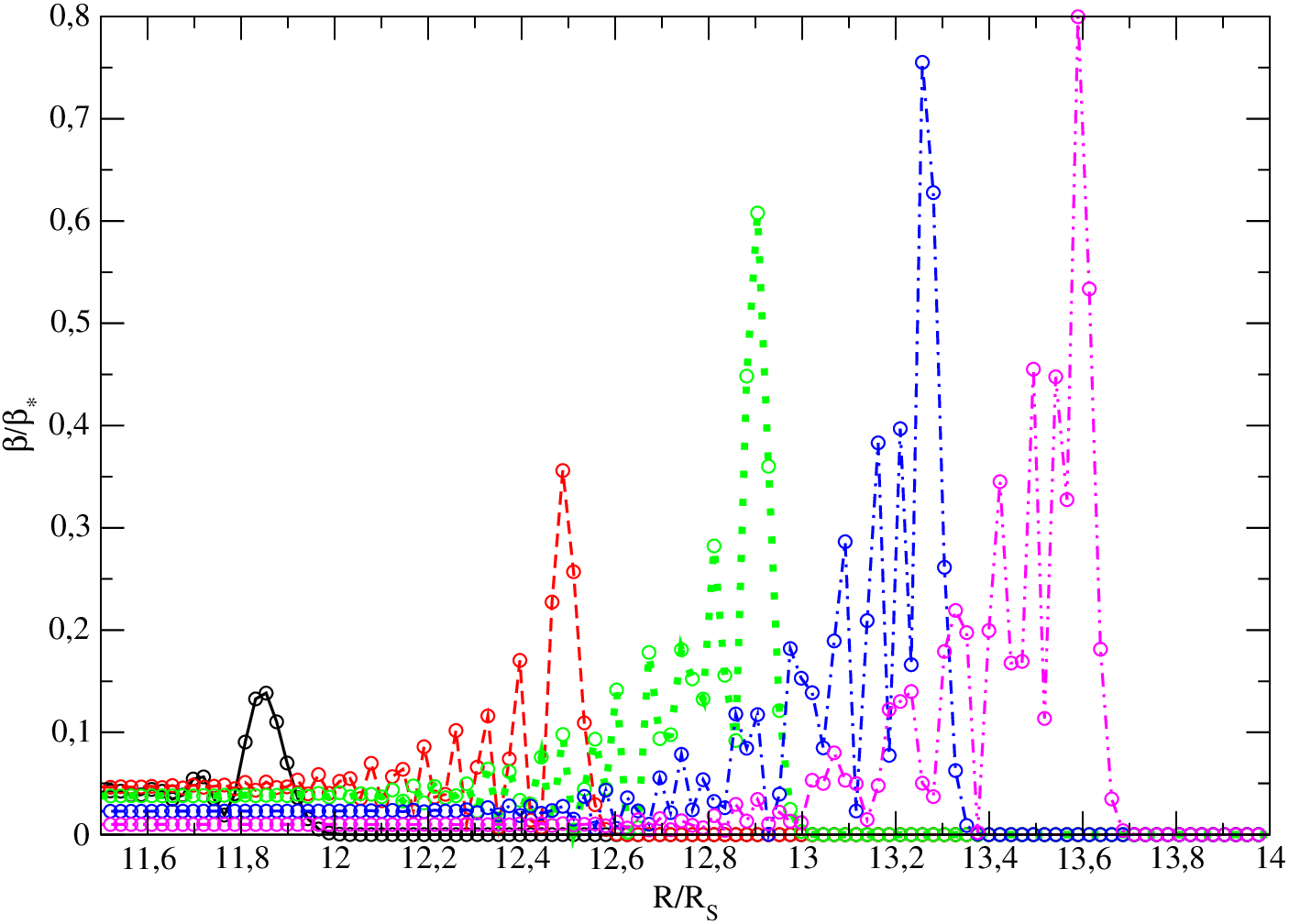}
\vspace{0cm}
\caption{The dependence of the ratio of the disc inclination $\beta$ to the stream inclination $\beta_*$ calculated for the $\alpha=0.01$ run at the same moments of time as those corresponding to the  curves plotted  in Figs. \ref{Fig1} and \ref{Fig2} is shown in the left panel . As for those Figs. solid, dashed, dotted, dot-dashed and dot-dot-dashed curves correspond to $t/P_{min}\approx 11,12,13,14,$ and $15,$
respectively. 
 The region in the neighbourhood of the propagation front is shown in greater detail  in the right panel. Symbols
indicate the locations of grid points.
}
\label{Fig9}
\vspace{0.5cm}
\end{figure}

In Fig \ref{Fig9} we show the dependence of the inclination angle $\beta$ on radius
$R$ calculated in a run with a small value of $\alpha=0.01$, for the same moments of
time as shown in Figs \ref{Fig1} and \ref{Fig2}. Note that in this run we set $\alpha_1=\alpha$
in (\ref{eqW}), begin  the calculation at $t=10P_{min}$ assuming that the disc is flat and lies in the equatorial plane
at that time. 
Note that because this is a linear response calculation, the calculated value of $\beta$ scales with $\beta_*.$ 
There are two distinctive features seem in the functional forms  of  the inclination angle.
Firstly, at a radii ranging from  $R \sim 11-14R_S,$ with larger radii corresponding to later times,
 we see very sharp changes of $\beta$ with typical amplitude increasing with time. 
 When $t=11P_{orb}$   $\beta$  has a resolved spike at   $R \sim 11.85R_S.$,  At later
times  short wavelength oscillations develop. Fig. \ref{Fig9} shows that the  wavelength of the  oscillations
attains  the order of grid size, and, therefore, at times larger than approximately $12P_{min}$ 
our grid  is not adequate to resolve the length scales associated  with the behaviour of our system.
We have checked
that when the grid size is reduced     
there  is always  a  time when the wavelength of
the oscillations gets comparable to that size.
 Also, one can check that, regardless of grid size and
 type of numerical scheme,  residuals in   a representation of the law of conservation of angular momentum (see
equation C5 in \cite{Zhu2014}) get progressively larger with time and eventually become too
large for the computation to be reliable.

There is a simple physical explanation for this behaviour of the system. As we have mentioned
above when $\alpha $ is sufficiently small, the  twist and tilt propagate with a speed equal to a
half  that of sound. At any moment of time a maximum of $\beta $ is approximately located at  the radius, $R_f$, 
where half of sound speed, $c_s$,  is equal to  the speed of outward moving material in  the hot phase 
which is also close to the propagation speed of the front, $v_f.$
Noting  that before $t/P_{min} \approx 15,$  $v_f$ is approximately constant we can assume 
that near and interior to $R_f$ the sound speed is approximately stationary in the frame moving with $v_f$
and such that  $c_s/2-v_f$ is positive.
On the other hand the sound speed decreases outwards. Therefore, the propagation speed of bending waves in the moving frame, $c_s/2-v_f$, becomes 
zero at $R_f$ and bending waves moving outwards from inner radii  accumulate there as their wavelengths arbitrarily shorten, the pile up eventually 
leading to a singularity  in the absence of a physical mechanism that allows disturbances induced by  the outgoing bending waves
to pass through the front. In that case numerical instability occurs.
Physically, the manifestation of  this singularity would be modified either by taking into account either non-linear terms  in
$\beta $  or including  higher order  terms in the expansion in powers of  $\delta$ that  has has been used in order to obtain equations (\ref{eq_B})
and (\ref{eqW}).
The latter approach make the  propagation of bending waves dispersive,  potentially  allowing 
leakage  through the critical point $R=R_f$. However, both approaches are technically complex
and  need to be studied  in depth  in three dimensions. 
Here, in order to resolve this difficulty, we  artificially increase
the coefficient $\alpha_1$ in (\ref{eqW}) in a region close to $R=R_f$.
This enables accumulating bending wave action to diffuse away
from $R=R_f$ at a rate governed by the magnitude of $\alpha_1$ thus avoiding the  production of  singular behaviour.
 From equation (\ref{e4}) it
follows that this term is in fact, of order $\delta^{2}$ with respect to the leading dissipative term $\propto \alpha$, and that making it artificially large doesn't spoil the 
dynamics of our system as long as scales much larger than $H$ are considered and 
$\delta_f=\delta (R_f)$ is not too large. Accordingly we expect our approach to give similar results to any procedure
that resolves the issue through diffusive effects limiting the development of small scales near $R=R_f.$

Typically, $\delta_f \sim 5\cdot 10^{-3}$ (see Fig. \ref{Fig1})  and, as seen from (\ref{e4})  when formally considering scales order of $R$ and, accordingly, $k \sim 1/R$ we can introduce very large values of $\alpha_1$: $\alpha_1 \sim < 10^5 \alpha $ without  disturbing the 
dynamics of our system on  these scales even without localising its application as indicated below.
 In our numerical work below we are going to consider
$$\alpha_1=\alpha_1(R=R_f)(1+\exp{-{(R-R_f)^2/( 4L^2_{\alpha_1})}}).$$
\noindent The spatial scale $L_{\alpha_1}$ is chosen to be the maximum
of  either $H(R_f)$ or the distance between neighbouring  grid points at $R=R_f$.
This will be done only for the runs with $\alpha=0.01$, which exhibit this type of instability.
 As  will be seen, introducing artificially large $\alpha_1$ can indeed regularise the behaviour of our system close to $R= R_f$ producing there a spike in $\beta$  with a finite amplitude
 \footnote{Clearly,  the amplitude of this spike has no direct physical origin, since  it is determined by the ad hoc imposition of a  value of $\alpha_1(R=R_f).$ 
In order to find a more realistic behaviour of $\beta $ 
close to the location of the front  one should invoke physically motivated methods of dealing with the  instability as discussed in the text.}
    We check below that making $\alpha_1$ ten times larger or smaller practically does not influence  our results,  apart from making the spike amplitude smaller or larger, respectively. 

Another feature  concerning functional forms  of $\beta $ and  $\gamma$ is that they are close to 
being  constant for $R < R_f$ during the stage when the front separating the hot phase from the cold phase ahead of it propagates outwards. 
An explanation for this easily follows from equations (\ref{e1}) and (\ref{e2}).
 Indeed, let us assume that both ${\cx B}$ and ${\cx W}$ vary in time on some time scale expressed in dimensionless units, $T_*,$ 
 and, accordingly, $\dot {\cx B} \sim {\cx B}/T_*$ and $\dot {\cx W} \sim {\cx W}/T_*$. 
 We also assume, as suggested by numerical calculations, that spatial scale of variation of ${\cx B}$ is $R$, while the corresponding  spatial scale of variation
  of ${\cx W}$, $L$, should be found from (\ref{e1}) and (\ref{e2}). 
  Retaining  only the terms containing  the highest order  spatial derivatives on the right hand sides 
   while  neglecting the contribution of the term  proportional to $\alpha_1$, we get from (\ref{e1}) 
   that $|{\cx B}| \sim T_{*}|{\cx W}|/(R^2L)$, while from (\ref{e2}) we get
   $|{\cx W}| \sim \delta^2 T_* |{\cx B}|$. 
   From these two expression we get $L \sim (T_*/T_s)^2R$, where $T_{s}=\delta^{-1}\Omega^{-1}$ is a characteristic time for a sound wave to propagate across the region of interest and we recall  that $\Omega =R^{-3/2}$ in our dimensionless  units.
    Thus, $L \gg R$ as long as $T_* \gg T_s$, being  the  condition that ${\cx W}$ should remain approximately constant
    over a scale $R$  which is satisfied in the inner region of the disc filled by a relatively hot gas. 
    This situation is similar to the known case of sufficiently thick tilted accretion discs in close  binary systems \citep[see][]{Larwood}.

\subsection{An ordinary  differential equation giving an approximate description of
the  time evolution of ${\cx W}$ for 
$R  < R_f$}\label{ODE}

That $\cx{W} $ is nearly constant inside $R_f$ during the stage of outward propagation of the  hot phase
allows us to reduce the evolution equations for ${\cx W}$ and ${\cx B}$ to a single first order equation using 
equation (\ref{e2}). 
To do this we multiply both parts of this equation $\xi R/\delta^2$ and integrate 
over a region from the marginally stable orbit to $R_f$. The surface terms may be shown to
be unimportant and, therefore, we obtain
\begin{equation}
\dot {\cx W}_0={1\over I_{1}}\left[2ia {G^2 M^2\over c^3}I_2 {\cx W}_0 + {\dot M_S R_S^{1/2} \over 2\pi}
({\cx W}_*-{\cx W}_0)\right], 
\label{e6}
\end{equation}
Here ${\cx W}_0$ is the value of ${\cx W}$ in the inner hot region
 which is assumed to depend only on time,
\begin{equation}
I_1=\int dR R^{3/2} \Sigma, \quad {\rm and}\quad  I_2=\int dR R^{-3/2}\Sigma, 
\label{e7}
\end{equation}

Here we have  used (\ref{eq9}), (\ref{eq11}) and  have temporarily restored physical units. The integration in (\ref{e7}) 
is formally performed from the radius of the  last stable orbit to $R=R_f$. The contribution of 
boundary terms associated with  both limits may be shown to be unimportant  since the surface density 
has its maximum near $R_S$ and the factor $D^{7/10}$ that scales the surface density vanishes at the last stable orbit
( see the discussion below equation  (\ref{eq2011c})) .

The expression (\ref{e6}), which can be regarded as an ordinary differential
equation for  ${\cx W}_0,$ reflects the law of conservation of angular momentum for the disc matter
with $R < R_f.$ This is affected only by the source produced by the stream and the gravito-magnetic torque. 
 We solve this equation  for ${\cx W}_0$
numerically and compare the result with results of solution of the full set (\ref{eq_B}) and
(\ref{eqW}) below.

\subsection{An approximate treatment of the transition region close to $R_f$}\label{simplefront}

Now let us consider the region close to $R_f$. In this region we assume that the aspect ratio $\delta $ as
well as variables ${\cx W}$ and ${\cx B}$ are stationary in the frame moving with front propagation speed
$v_f$.  Accordingly the  time derivatives in (\ref{e1}) and (\ref{e2}) can be replaced  by $-v_f \partial / 
\partial R$ with the consequence that these equations become ordinary differential equations. We introduce 
a dimensionless variable $y=1-R/R_f$ and consider only the region for which
 $y \ll 1.$  All coefficients in (\ref{e1}) and (\ref{e2}) apart from $\delta $
will be assumed to vary slowly compared  to ${\cx B}$ and ${\cx W}$  so that  they can be replaced  constants 
equal to their  value evaluated at $R=R_f.$ 
 Note that this assumption, made for simplicity,  is rather crude since $\Sigma $ has significant  variation in  this  region.
In this context we remark that the purpose of this  Section is to construct an approximate
theory  that exhibits the main features of the transition region rather that a precise quantitative representation. 
In this spirit we also neglect terms proportional to $\Omega {\cx B}$ on the right hand side of (\ref{e1})
and  last two terms on the right hand side  of (\ref{e2})  due to Lense-Thirring precession and 
that the stream respectively. 

Equations (\ref{e1}) and (\ref{e2}) can then be integrated once immediately.
The resulting form  of (\ref{e1}) can then be used to express ${\cx B}$ in terms of ${\cx W}$  and the result used in
 (\ref{e2}) after integration. In this way find that ${\cx W}$ satisfies 
\begin{equation}
\delta^2\alpha_1\tilde v {d {\cx W}\over d y}+\left(\delta^2 -{\tilde v}^2\right){\cx W}=C_1+C_2\delta^2,
\label{e8}
\end{equation} 
where $\tilde v =2v_f/v_K=2\sqrt{R_f}v_f$  with $v_{K}=R_f^{-1/2}$ being  the Keplerian speed of circular motion expressed in our dimensionless
units  and $C_1$ and $C_2$
are constants of integration. Note that we assume that the front propagates with the sound speed 
at $R=R_f$ so that we have $\tilde v=\delta(R_f)$.

When the term proportional to $\alpha_1$  is negligible, as expected significantly inward of the front, we get immediately from (\ref{e8})
\begin{equation}
{\cx W}={C_1+C_2\delta^2\over \delta^2 - {\tilde v}^2}.
\label{e9}
\end{equation} 
When $\delta \gg \tilde v$ the solution (\ref{e9}) must coincide with the solution,  ${\cx W}_0,$ given by   (\ref{e6}) that is applicable to the interior region.  
 This leads to the requirement that $C_1=0$ and $C_2={\cx W}_0$. 

An analytic solution of (\ref{e8}) can be found very close to $R_f$, by  expanding  $\delta $ in  a Taylor series
in $y$ up to the first order  and then setting  $\delta^2 -{\tilde v}^2= 2\delta\delta^{\prime}y,$  where the $\prime$ denotes the derivative with respect to
$y$ evaluated at $y=0,$ and $\delta^2= {\tilde v}^2$ elsewhere 
in (\ref{e8}) which then gives
\begin{equation}
\alpha_1 {\tilde v}^2 {d {\cx W}\over dy}+2\delta^{\prime} y {\cx W}={\tilde v}{\cx W}_0.
\label{e10}
\end{equation} 
A  solution to (\ref{e10}) for $y > 0,$ can be expressed in terms of the  Dawson function \\\
$Daw(z)=\exp(-z^2)\int_0^z dt \exp(t^2)$. It takes  the form
\begin{equation}
{\cx W}=\frac{1}{\sqrt{{\alpha_1 \delta^{\prime} }}}\left({\delta \over \tilde v}\right)^2{\cx W}_0Daw(z), \quad {\rm with}\quad z=\sqrt{\frac{\delta^{\prime}}{\alpha_1  {\tilde v}^2}} y,
\label{e11}
\end{equation}  
where we remark  that $\delta^{\prime} > 0.$  Note also  the factor $(\delta/  \tilde v)^2$ inserted  in 
(\ref{e11}). It is close to one near the critical point, but is needed to smoothly match the solution
to ${\cx W}_0$ at smaller radii. In addition we  assume  that the disc has
zero inclination upfront of the critical point which is satisfied by (\ref{e11}) as  $D(z=0)=0.$  

One can can see that (\ref{e11}) is approximately equal to (\ref{e9}) in the asymptotic limit $z \gg 1$, but with $y \ll 1$
\footnote{Note that in order to have such a limit we need to have $\alpha_1{\tilde v}^2/\delta^{\prime} \ll 1$, which is easily satisfied.}
It is possible to obtain an approximate form of ${\cx W}$, which is always close to (\ref{e9}) when $z \gg 1$ and always
close to (\ref{e11}) when $y \ll 1$. It can be done through redefinition of the variable $z$ through
\begin{equation}
z={1\over 2}{\delta^2 -{\tilde v}^2\over \sqrt{\alpha_1\delta^{\prime} \tilde v^4}}.
\label{e12}
\end{equation}
It is straightforward to see that when $y \ll 1$ $z$ is reduces to the previous definition. That ${\cx W}$ given by
(\ref{e11}) and (\ref{e12})  reduces to the form (\ref{e9}) follows from the asymptotic form of $D(z)$: $D(z)\approx 1/(2z)$ 
when $z \gg 1$. 
\begin{figure}
\begin{center}
\vspace{10cm}\hspace{-10cm}\includegraphics{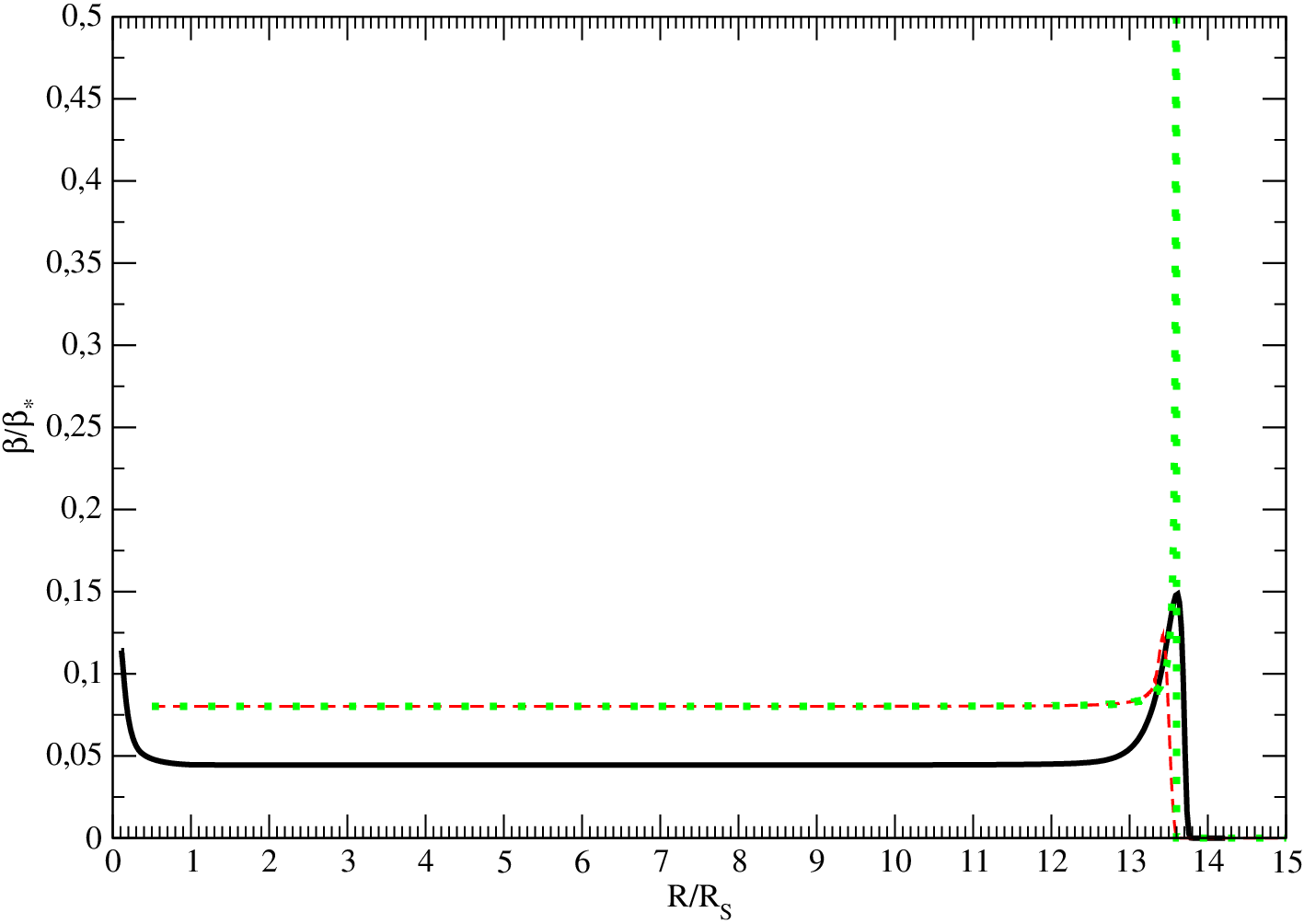}
\end{center}
\vspace{0cm}
\caption{A comparison of the dependence of $\beta$ on $R$ in a numerical calculation (solid curve) and the semi-analytic model
developed in this Section (dashed and dotted curves). Parameters and initial conditions of the run are the same as used to 
obtain the curves shown in Figs \ref{Fig9} 
except that  we now set $\alpha_1(R=R_f)=10^3\alpha$.}
\label{Fig11}
\vspace{-0.5cm}
\end{figure}

We compare our semi-analytic model with numerical data in Fig. \ref{Fig11}, where we show dependencies of $\beta $ on $R$ 
at the time $t=15P_{min}.$
in a run with $\alpha=0.01$ \footnote{Let us recall that the actual value of $\alpha$ used in the numerical solution for $\beta$
is $(2/3)0.01\approx 0.0067$ due to the redefinition of $\alpha $ discussed above.}  The solid curve represents the fully numerical
calculation. The dashed curve is obtained using the expressions (\ref{e11}) and (\ref{e12}), while to plot the dotted curve we use
the expression (\ref{e9}). In both cases ${\cx W}_0$ is calculated by solving  equation (\ref{e6}) numerically. We see that 
$\beta_0\equiv |{\cx W}_0|$ overestimates the actual value of $\beta $ by about 60 per cent. Also, although a typical width of the semi-analytic curves is similar to that of numerical one, the maximum value of $\beta $ is roughly 20 percent smaller  with a much sharper variation of $\beta $ near the maximum while the total range of $\beta$ is a factor of three smaller. We believe that all of these defects arise through the  crudeness of our semi-analytic model 
and  the situation could be improved upon in a more accurate treatment.

\section{Numerical results for the evolution of the disc inclination and twist} \label{numres}
 
In this Section we  discuss  numerical solutions of equations (\ref{eq_B}) and (\ref{eqW}) in detail as 
well as their comparison with the semi-analytic model developed in the previous Section and the quasi-stationary model developed
in \cite{Xiang2016}. For details of the numerical scheme  we used see Appendix \ref{scheme}.
Our numerical runs are distinguished by values of the viscosity parameter, $\alpha$, rotation parameter $a$,
initial conditions, which  correspond to either a flat disc lying in the equatorial plane or in the plane associated with
the stream, and time of the commencement of  the simulations.
 Let us recall  that on account of differences in the procedure for reducing the problem to a one dimensional one,
 in equations (\ref{eq_B}) and (\ref{eqW}) we use
values of $\delta $ and $\alpha $ smaller than those used in the equations describing the evolution of background quantities 
by factors $1/\sqrt{2}$ and $2/3$, respectively. 
Nonetheless, we label cases corresponding to different $\alpha$
with  its value used in the calculation of the  background quantities, and only two such values, $\alpha=0.01$ and $\alpha=0.1$
are considered below, we recall that $\alpha=0.1$ and $\alpha=0.3$ were considered in \citet{Xiang2016}. 
We consider the cases corresponding to $a=1$, initial time of computation, $t_{in}=10P_{min},$ and initial disc plane coinciding
with the black hole equatorial plane and, accordingly, $\beta(t=t_{in})=0$, in more detail than the others. These cases are referred
hereafter to as  standard.

\subsection{The dependence of the  inclination angle evaluated at the stream impact position on time}\label{streaminc}

An important quantity, which  allows us to make a comparison of our fully numerical results with those  obtained 
using different semi-analytic schemes is the  time dependence of the inclination angle $\beta $ evaluated at the stream impact 
radius $R_S$. This quantity also characterises the  dependence of typical disc inclinations on time. Let us stress that it 
is not necessarily the largest value of $\beta $ at a particular time,  the latter could be situated at larger or smaller
radii depending on the parameters of a particular calculation and time.   

\subsubsection{The case $\alpha=0.01$}

\begin{figure}
\begin{center}
\vspace{10cm}\hspace{-10cm}\includegraphics{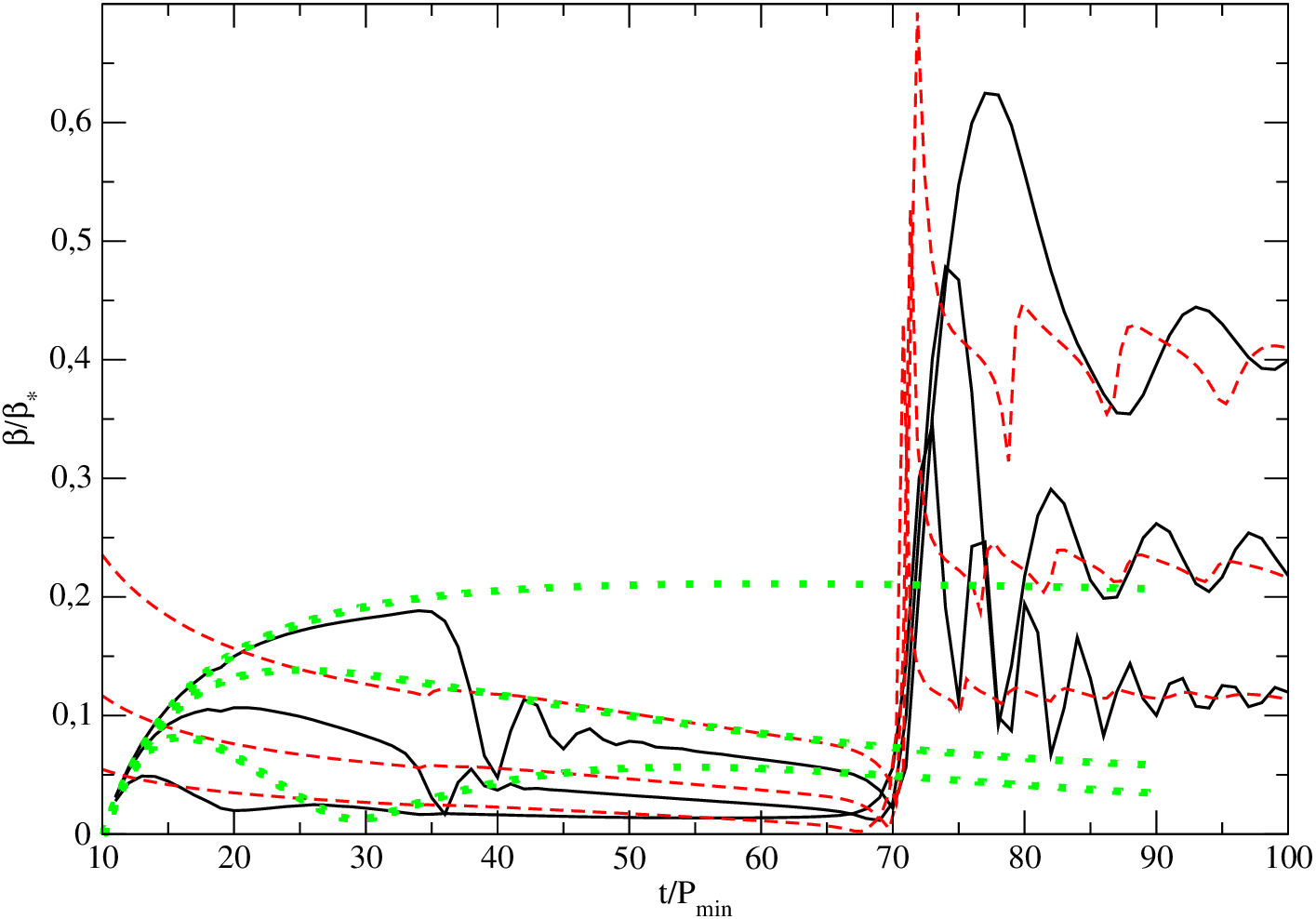}
\end{center}
\vspace{0cm}
\caption{The dependence of inclination angle $\beta $ evaluated at the stream impact position on time. 
 Cases with prograde black hole rotation, $a > 0$, are shown for $a= 0.25, 0.5$ and $1.0.$ 
The solid curves correspond to the numerical solution of equations (\ref{eq_B}) and (\ref{eqW}), the dashed curves are calculated according to the quasi-static approach described in \citet{Xiang2016}
 and the dotted curves are obtained by numerical solution of 
equation (\ref{e6}). For curves of a given type, larger values of $\beta$ correspond to smaller values of $|a|.$
}
\label{Fig12}
\vspace{-0.5cm}
\end{figure}

\begin{figure}
\begin{center}
\vspace{10cm}\hspace{-10cm}\includegraphics{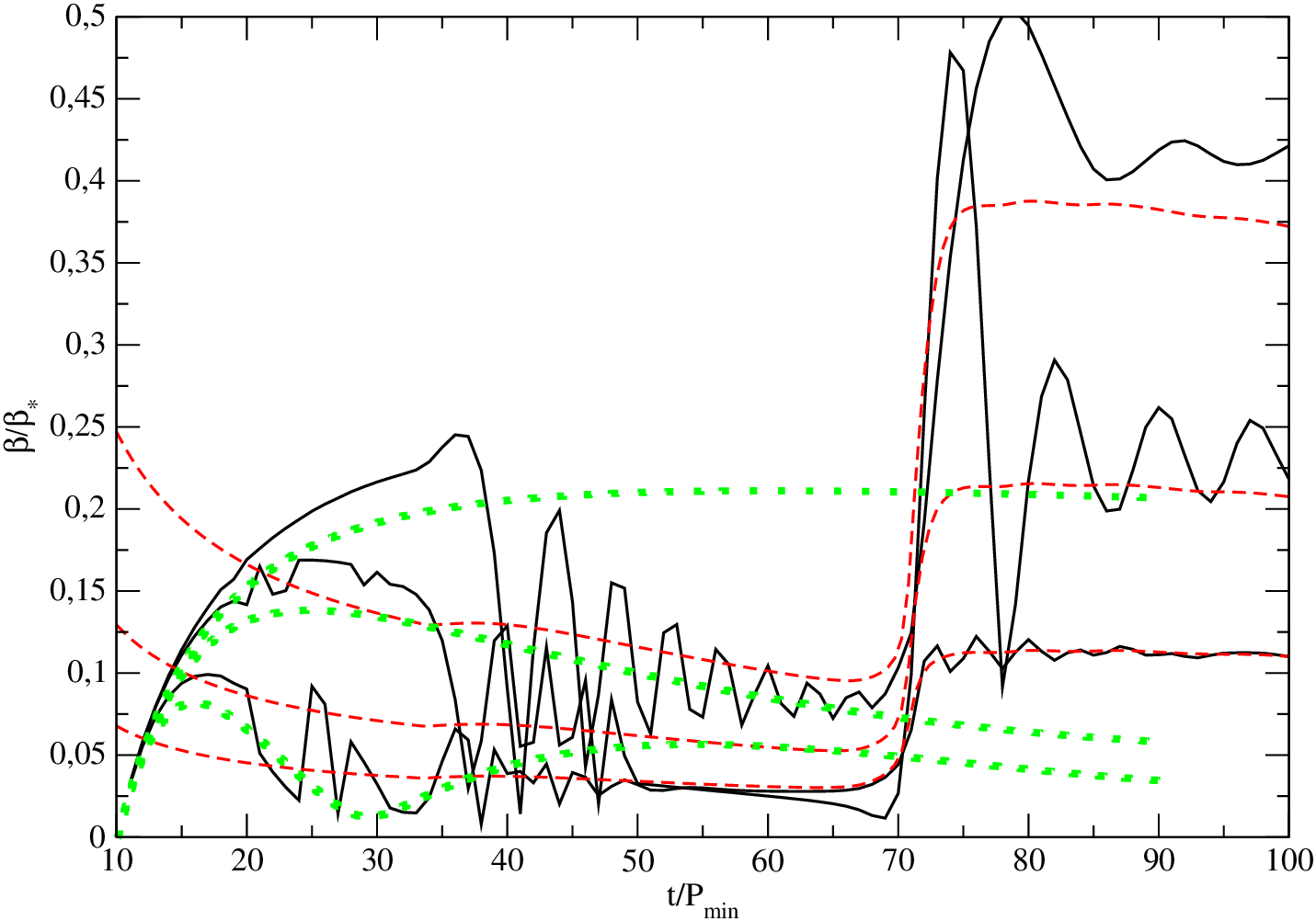}
\end{center}
\vspace{0cm}
\caption{Same as Fig. \ref{Fig12}, but for retrograde black hole rotations, $a < 0$.
When comparing with Fig. \ref{Fig12}, corresponding curves are such that 
$a \rightarrow -a.$
\label{Fig13}}
\vspace{-0.5cm}
\end{figure}

In Figs. \ref{Fig12} and \ref{Fig13} we show the dependence of $\beta (R_{S})$ on time for prograde and retrograde black
hole rotation, respectively with absolute values of the rotational parameter $|a|=0.25, 0.5$ and $1.0$. Other parameters of the 
calculations have standard values.
 As seen from these Figs. values of the  inclination get  larger when the absolute value 
of $a$ decreases as expected. A sharp rise of $\beta $ at $t/P_{min} \approx 70$ is associated with the development of
thermal instability as explained in \cite{Xiang2016}. It is also evident from these Figs. that our time dependent model described 
by equation (\ref{e6}) adequately describes the numerical solutions at relatively early times, $t/P_{min} < 15-35$, 
being significantly better for small, $|a|,$ while
the quasi-static solutions are close to the numerical ones at sufficiently late times.

\subsubsection{The case $\alpha=0.1$}

\begin{figure}
\begin{center}
\vspace{10cm}\hspace{-10cm}\includegraphics{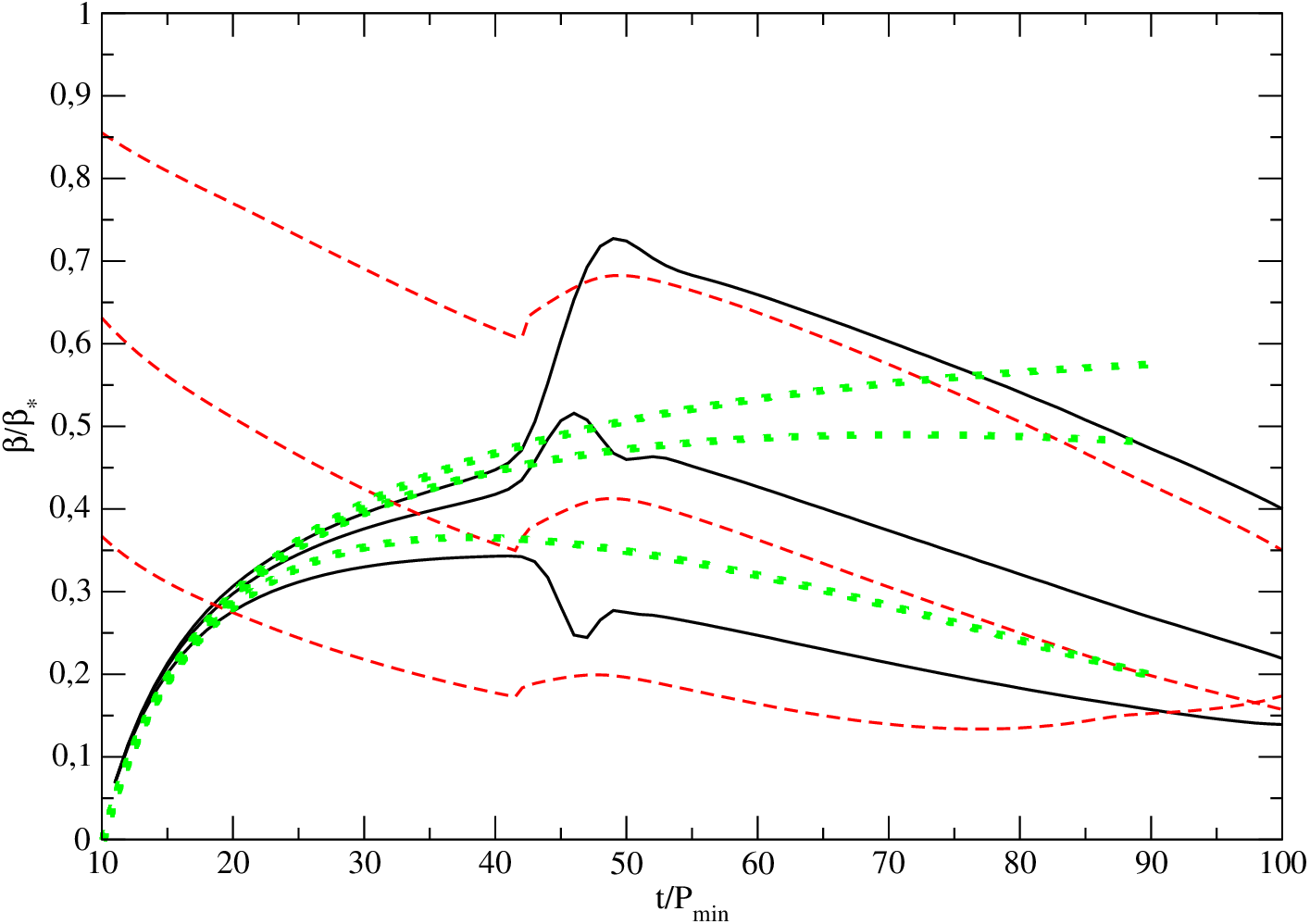}
\end{center}
\vspace{0cm}
\caption{Same as Fig. \ref{Fig12}, but for $\alpha=0.1$. Solid curves are obtained numerically, while the dashed ones 
are  calculated in framework of the quasi-static approach. The dotted curve represents calculations based
on the solution of equation (\ref{e6})  for the case $\alpha=0.01$. The case of $a > 0$ is shown.
\label{Fig14}}
\vspace{-0.5cm}
\end{figure}

\begin{figure}
\begin{center}
\vspace{10cm}\hspace{-10cm}\includegraphics{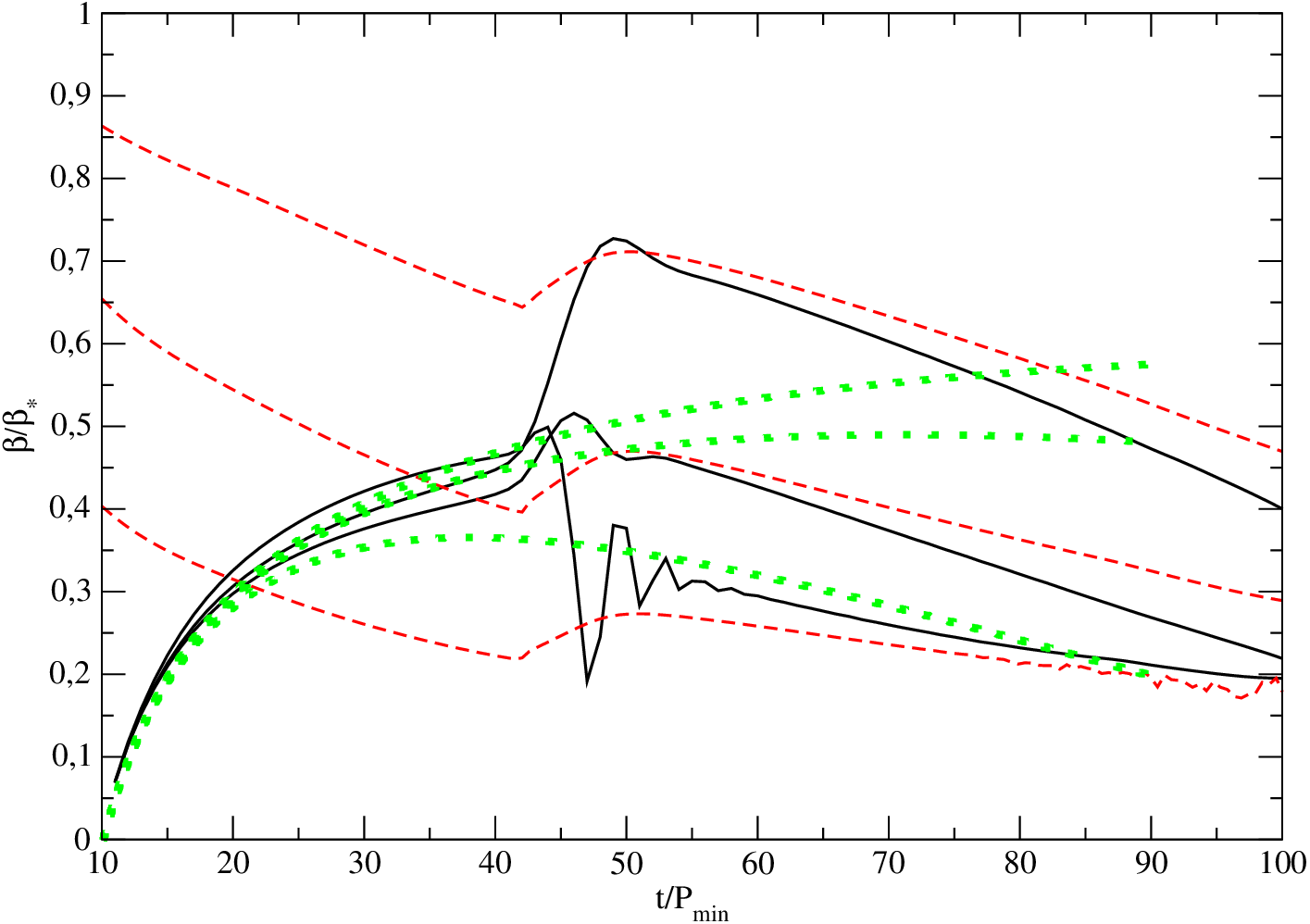}
\end{center}
\vspace{0cm}
\caption{Same as Fig. \ref{Fig14}, but for $a < 0$.}
\label{Fig15}
\vspace{-0.5cm}
\end{figure}

Figs \ref{Fig14} and \ref{Fig15} are analogous to Fig. \ref{Fig12} and \ref{Fig13}, but calculated for larger $\alpha=0.1.$
Since in this case there is no outward propagation of  a hot front during the time span of the simulations 
so that the  twist and tilt dynamics  is  for the most part determined by
viscosity rather than effects due to  pressure  and advection, we do not show results based on equation (\ref{e6}). 
From these Figs.
it follows that inclinations are, in general, larger than those corresponding to the case with $\alpha=0.01$ and 
that our quasi-static model describes the numerical results quite well at late times.

\subsubsection{The effect of changing the  initial conditions and varying $\alpha_1$}

\begin{figure}
\begin{center}
\vspace{10cm}\hspace{-10cm}\includegraphics{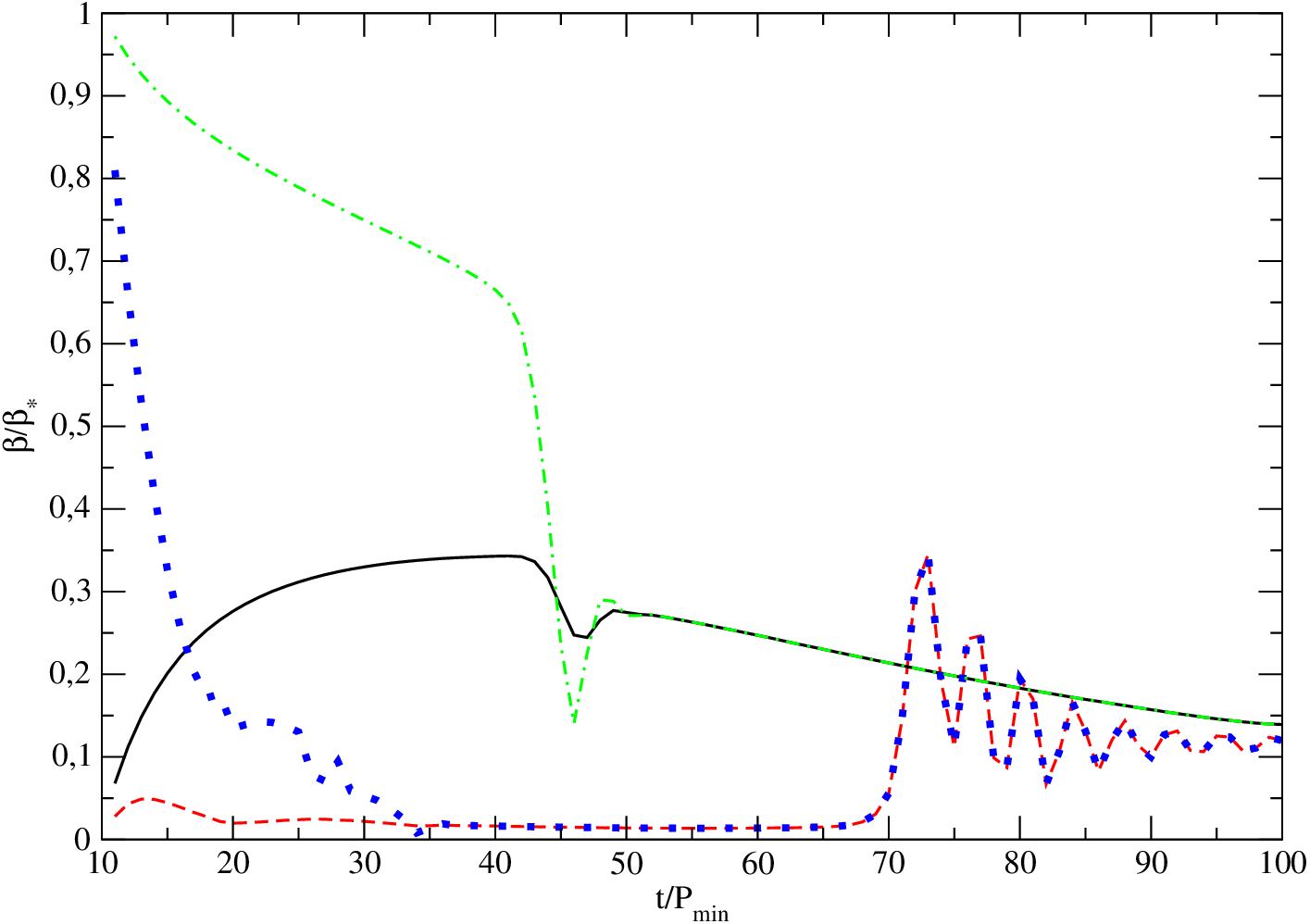}
\end{center}
\vspace{0cm}
\caption{  The evolution of $\beta$ at the stream impact position in  our standard cases is  compared  with that  obtained by changing 
 the initial value of $\beta$. Solid and dashed curves are for  our standard cases with
 $\alpha=0.1$ and $0.01$, respectively,
 while dotted and dot dashed curves correspond, respectively, to simulations with the same values of $\alpha $,
 but set  the  initial disc inclination to be equal
 to that of stream,( $\beta(t_{in})=\beta_{*}$).
At late times curves corresponding to different values of $\beta_{in}$ but which have the same value of $\alpha$
 overlap completely}
\label{Fig16}
\vspace{-0.5cm}
\end{figure}

\begin{figure}
\begin{center}
\vspace{10cm}\hspace{-10cm}\includegraphics{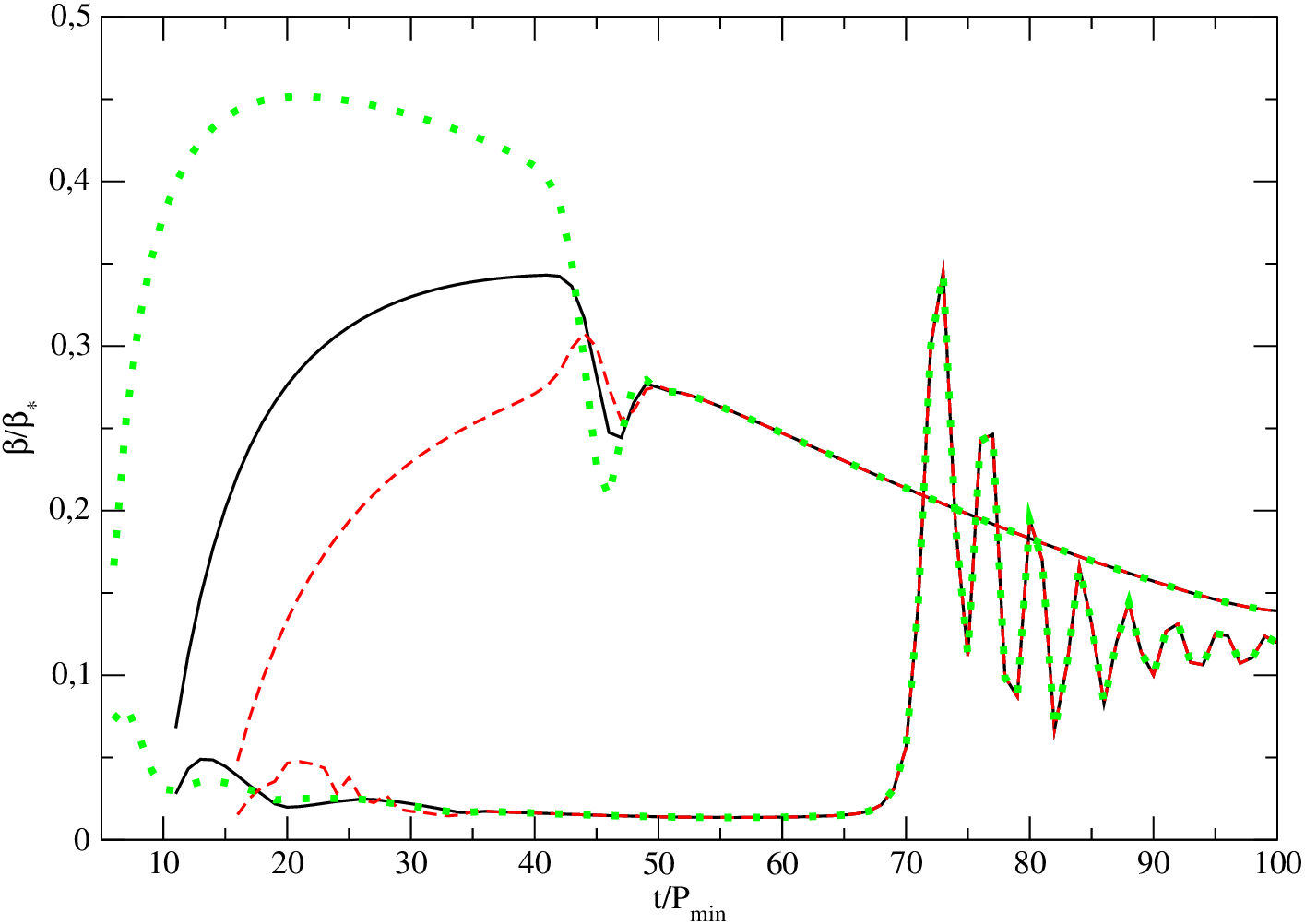}
\end{center}
\vspace{0cm}
\caption{Same as Fig. \ref{Fig17}, but now the  comparison is between  our standard cases 
and simulations that differ by the choice  of commencement time, $t_{in}$.
 The curves of the same style 
  with larger (smaller) values of $\beta $ at small times correspond to $\alpha=0.1$ ($\alpha=0.01$).
 Solid curves are for the  standard cases, while dashed and dotted curves  are calculated for $t_{in}=15P_{min}$ and
$t_{in}=5P_{min}$, respectively.}
\label{Fig17}
\vspace{-0.5cm}
\end{figure}

\begin{figure}
\begin{center}
\vspace{10cm}\hspace{-10cm}\includegraphics{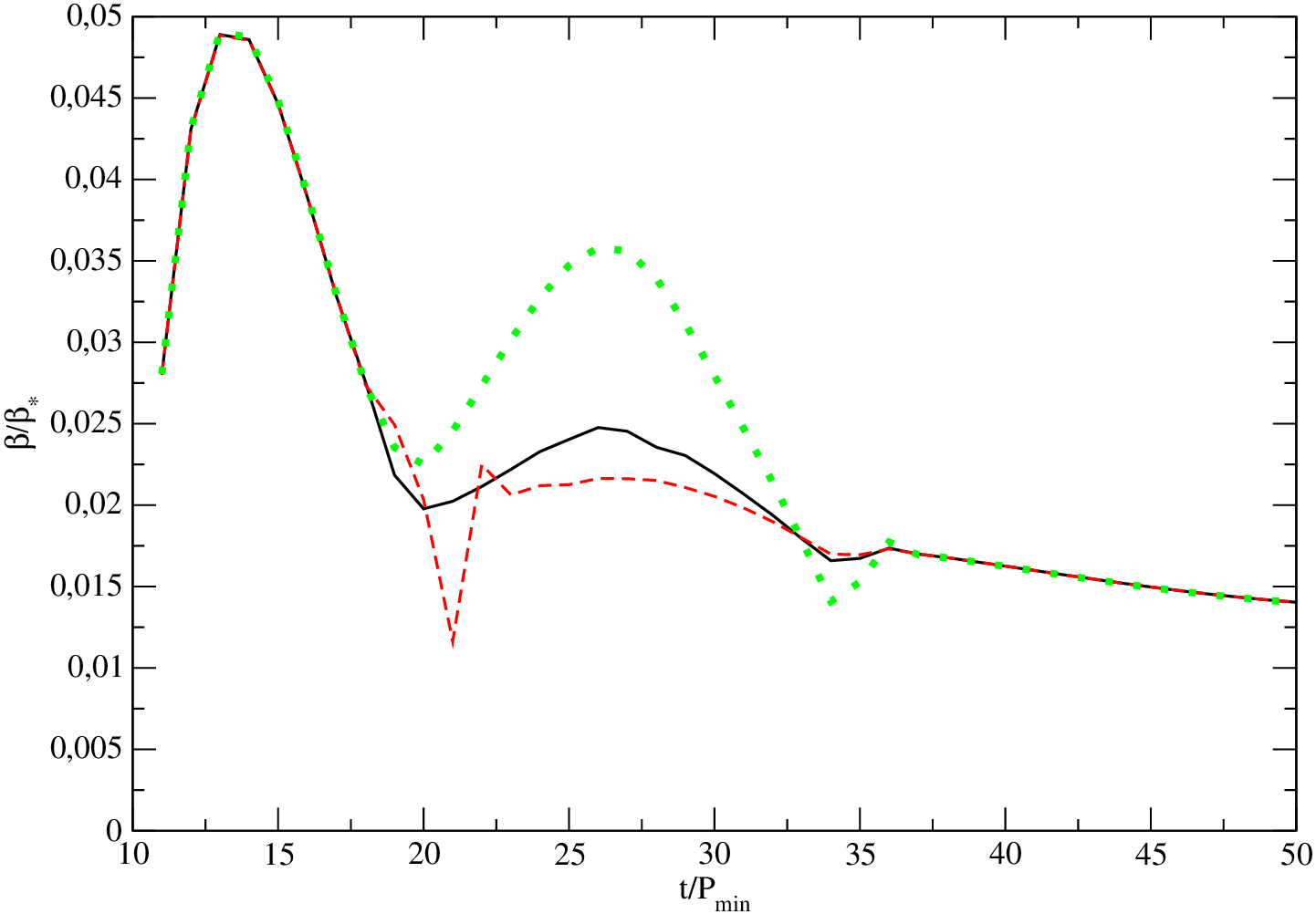}
\end{center}
\vspace{0cm}
\caption{Same as Fig. \ref{Fig17}, but now  curves  of different style 
correspond to different values of $\alpha_1$. Solid, dashed 
and dotted curves respectively represent  cases with
 the standard $\alpha_1=1000\alpha$, $\alpha_1=100\alpha$ and $\alpha_1=10000\alpha$,
respectively. Other parameters have standard values.
 Note that we show only $t/P_{min} < 50$, since the curves  completely overlap
at  later times.}
\label{Fig17a}
\vspace{-0.5cm}
\end{figure}

Figs. \ref{Fig16}-\ref{Fig17a} represent results obtained when varying  different parameters that specify a simulation.
In Fig. \ref{Fig16} we  change the initial inclination of the disc to $\beta_*.$ 
In Fig. \ref{Fig17} dashed and dotted curves
correspond to the  time that the  computation was commenced  being changed to $t_{in}=15$ and $5P_{min},$ respectively. 
In \ref{Fig17a} we illustrate  results obtained by changing  the value of $\alpha_1(R=R_f)$ to $100\alpha$ and $10000\alpha$ 
with respectively  dashed and dotted curves. The
value of  $\alpha$ is $0.01.$
Note that for each of the above  cases  only one parameter of the problem  is varied while keeping the others equal to their
standard values. 

As seen from these Figs. apart from when the  initial value of  $\beta $  is changed, variations of different parameters lead to only
modest changes of the results with,  in particular,  no noticeable deviations  at later times.
 The change of initial $\beta$ to $\beta_*$ leads to larger values of inclinations at relatively
early times, but the curves converge for  $t/P_{min} > 35$ when  $\alpha=0.01$ and for  $t/P_{min} > 45$ when  $\alpha=0.1$.

\subsubsection{The dependence of inclination angle on radius at specified  moments of time}

We now  illustrate the  dependence of $\beta $ on $R$ at  different moments of time. In the low viscosity 
case we show the curves calculated at 'early'   times,  $t/P_{min}=11,12,13,14$ and $15$ corresponding to 
the situation when the hot phase propagates outwards and profiles of $\beta$ are expected to be nearly uniform at small
radii. In addition  the form of the inclination is illustrated  at 'late'  times  $t/P_{min}=40,60,80$ 
and $100$ when the disc is expected to relax to its quasi-stationary shape. In the high viscosity case only the curves
corresponding to the same 'late' times are shown.

\subsubsection{The case $\alpha=0.01$}

\begin{figure}
\begin{center}
\vspace{10cm}\hspace{-10cm}\includegraphics{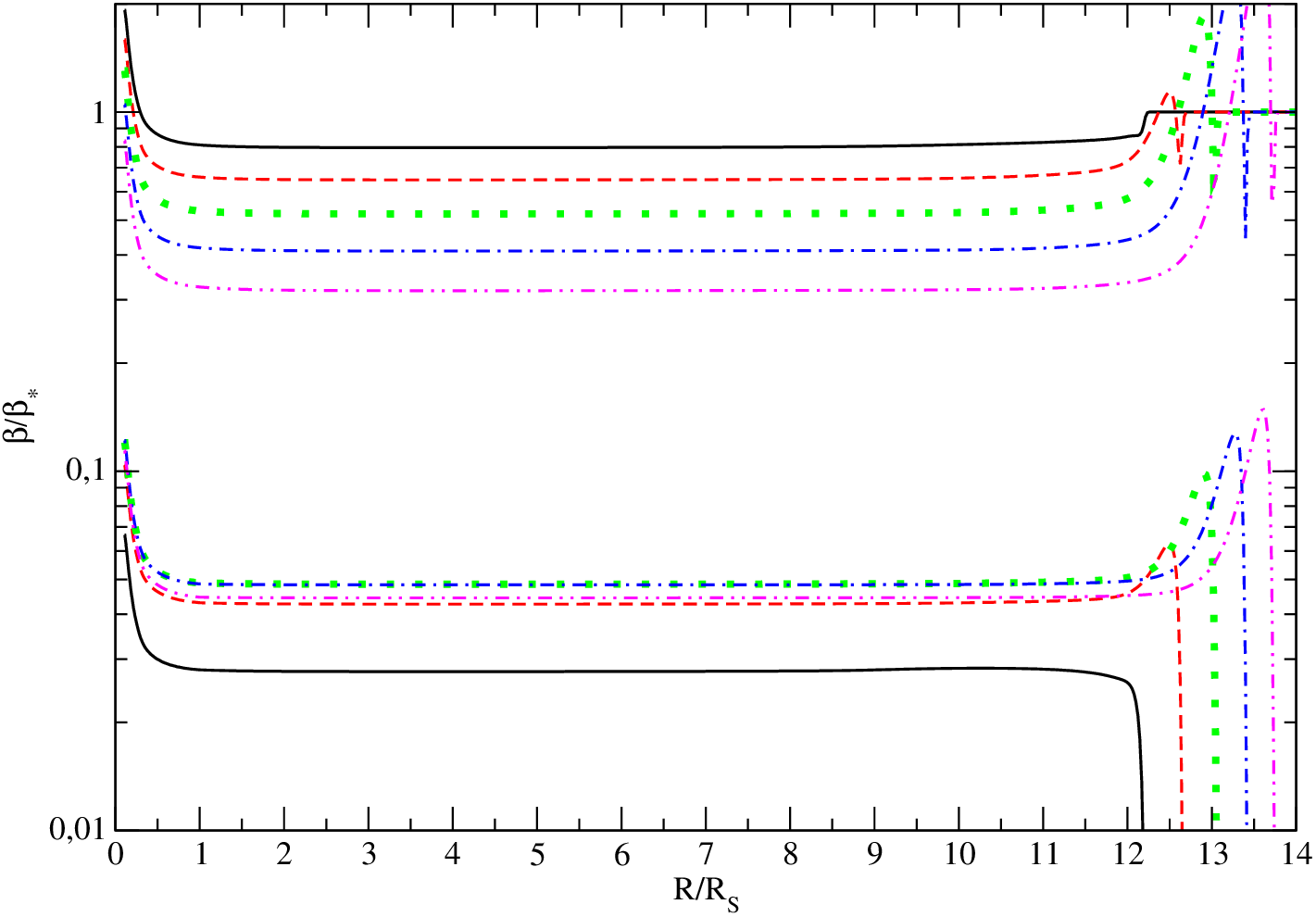}
\end{center}
\vspace{0cm}
\caption{Dependence of $\beta $ on $R$ calculated  at relatively early moments of time, 
when an  inner hot region  is present in the disc and $\beta $ is nearly constant in this region.
 Curves of the same type taking on  smaller and larger values at a given time 
 correspond to the standard case and the case differing from the standard one by setting
 $\beta (t_{in})=\beta_{*}$, respectively. 
Solid, dashed, dotted, dot dashed and dot dot dashed curves are calculated with $t/P_{min} =11,12,13,14$ and $15,$ respectively.}
\label{Fig18}
\vspace{-0.5cm}
\end{figure}

Fig. \ref{Fig18} shows profiles of $\beta $ at the  the 'early' times. Curves of a given  type  that take on larger
values at a given time  represent the calculation with $\beta (t_{in})=\beta_*$, all other parameters are standard. 
We see that the disc 
behaves in the expected manner, with inclinations almost independent of radius inside the front region and a spike in this
region. Note that the  amplitudes of the spikes  are determined by the value of $\alpha_1(R=R_f)$ and so may not be realistic.
We recall that this value does not affect the flow elsewhere (see Section \ref{propfront}).

\noindent Also note that values of the inclination grow slightly with decreasing  radius. This reflects the fact that a low viscosity
twisted disc does not align with the black hole equatorial plane when $a > 0$, see e.g. \cite{Iva1997}. We have checked that
the prograde case with $a < 0$ gives similar curves, but with inclinations tending to  alignment at small radii as expected.

\begin{figure}
\begin{center}
\vspace{10cm}\hspace{-10cm}\includegraphics{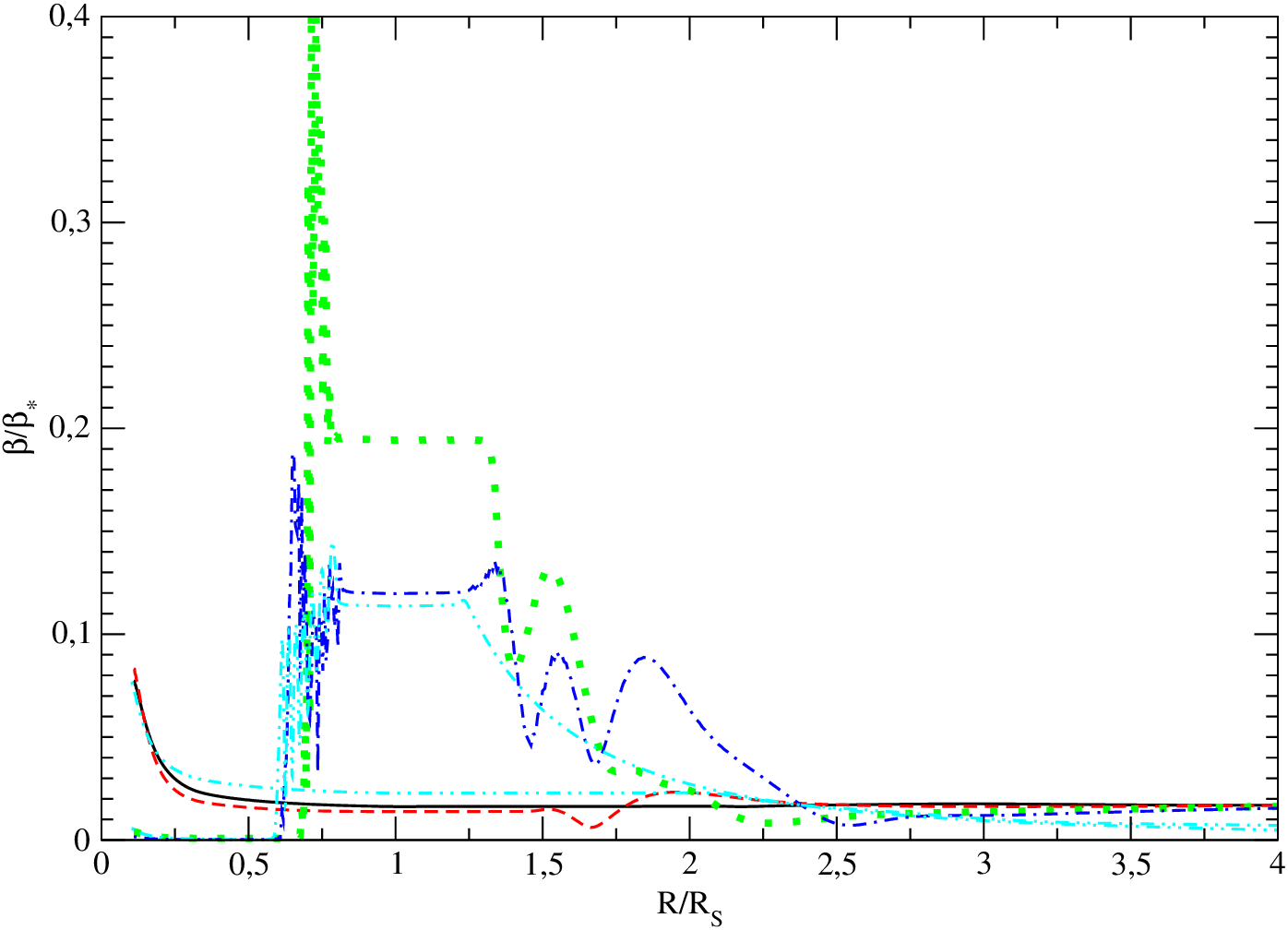}
\end{center}
\vspace{0cm}
\caption{Same as Fig \ref{Fig18}, but for the standard case and later moments of time. Solid, dashed, dotted and dot dashed curves correspond to $t/P_{min}=40,60,80$ and $100$, respectively. We also show the results of calculations based on the
 quasi-static approach by two dot dot dashed lines, with the curve for which $\beta$ is  nearly constant  for  $R/R_S \sim > 1$
 corresponding to $t=40P_{min}$ and the one showing much larger variations  of $\beta $  corresponding to $t=100P_{min}$.}
\label{Fig19}
\vspace{-0.5cm}
\end{figure}

\begin{figure}
\begin{center}
\vspace{10cm}\hspace{-10cm}\includegraphics{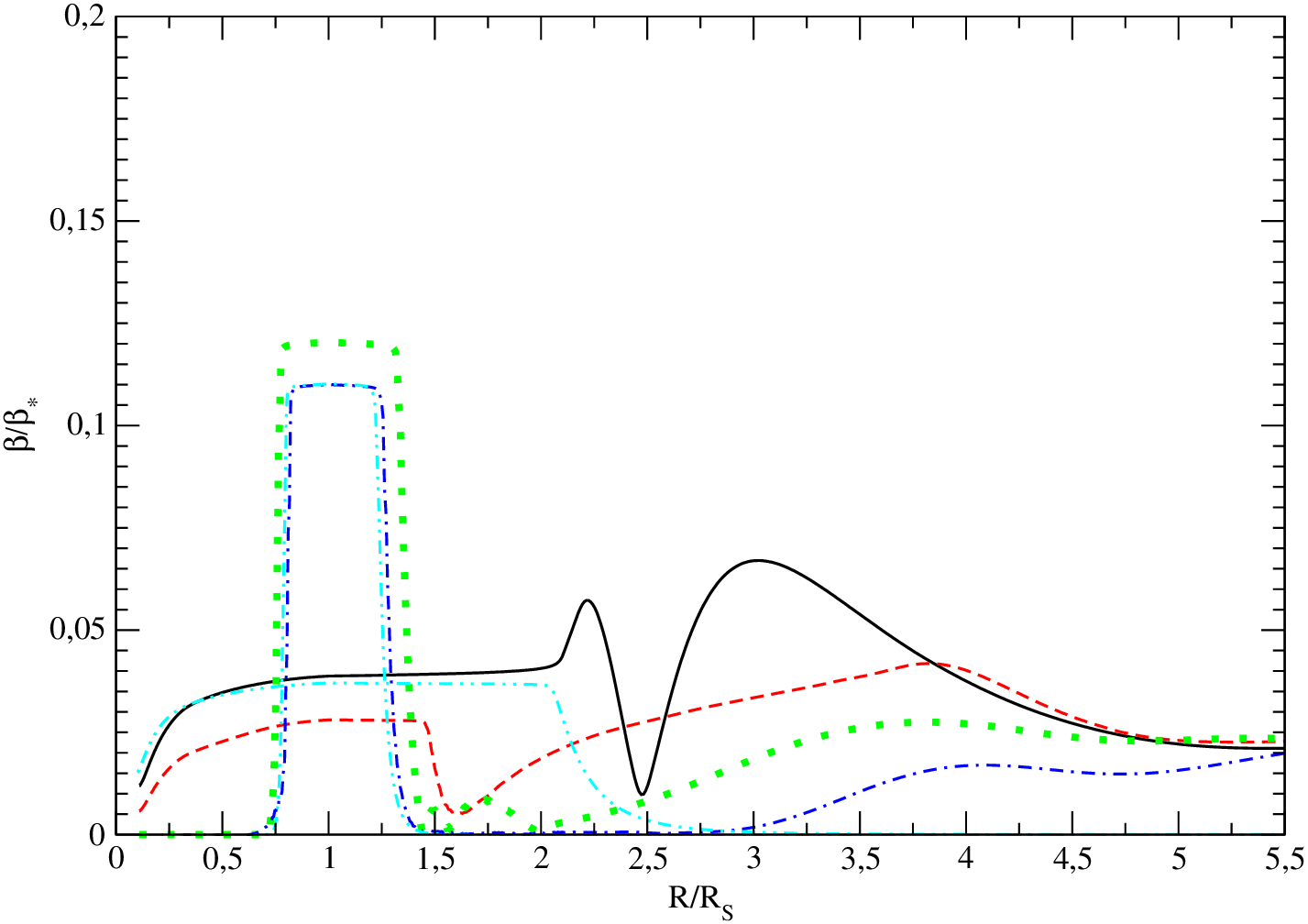}
\end{center}
\vspace{0cm}
\caption{Same as Fig. \ref{Fig19}, but for retrograde black hole rotation with $a=-1$. The dot dot dashed curve with larger values of $\beta $ at $R/R_S \sim $ is for $t=100P_{min}$ 
(note that it almost coincides with the corresponding  numerical 
curve), while  the other curve of the same type is for $t=40P_{min}$.}
\label{Fig20}
\vspace{-0.5cm}
\end{figure}

Figs \ref{Fig19} and \ref{Fig20} show  inclination profiles at  'late'  times. 
In addition we present 
two profiles calculated in the framework of our quasi-stationary approach when $t/P_{min}=40$ and $100$ as dot dot dashed curves.
At times $t/P_{min} \sim 100$ the the disc inclinations are large only in a region $R \sim R_{S}$ 
with the result based on the  quasi-stationary approximation being very close to the fully numerical one.  Note
that, as seen in  Fig. \ref{Fig19}, in the prograde case we have very sharp oscillations of inclination 
angle at $R/R_S \sim 0.75$. This is due to the fact, that a low viscosity twisted disc does not align with
the equatorial plane at small radii, instead producing a standing bending wave, \citep[see][]{Iva1997,Nea2015}. However,
the disc behaviour at close to this radius is, most probably, unphysical, since very sharp changes of $\beta$ 
are likely to result in Kelvin - Helmholtz like  instabilities and possibly an  increase in effective viscosity due to non-linear effects associated with their development.

\subsubsection{The case $\alpha=0.1$}

\begin{figure}
\begin{center}
\vspace{10cm}\hspace{-10cm}\includegraphics{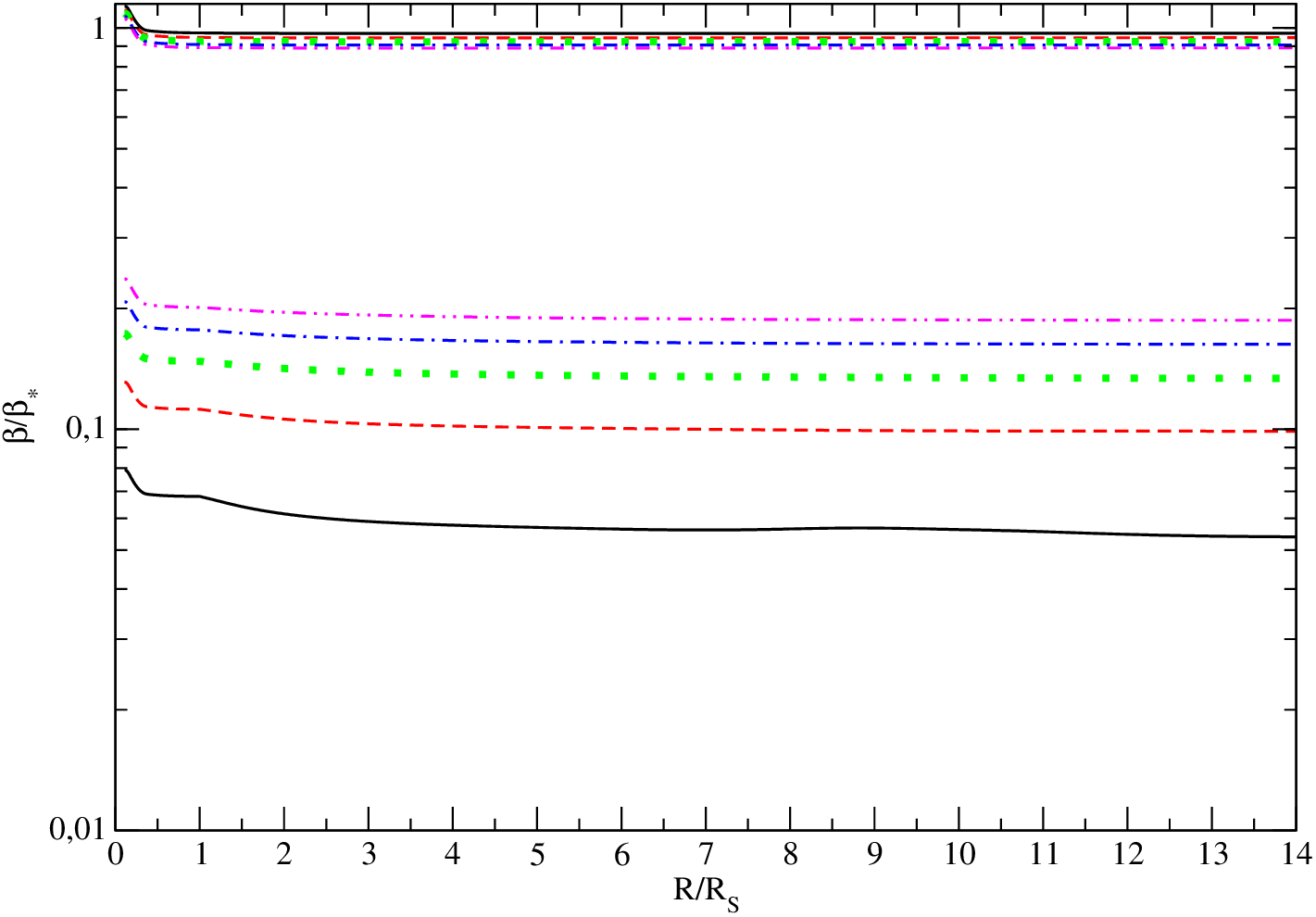}
\end{center}
\vspace{0cm}
\caption{Same as Fig \ref{Fig18}, but for $\alpha=0.1$.}
\label{Fig19a}
\vspace{-0.5cm}
\end{figure}

\begin{figure}
\begin{center}
\vspace{10cm}\hspace{-10cm}\includegraphics{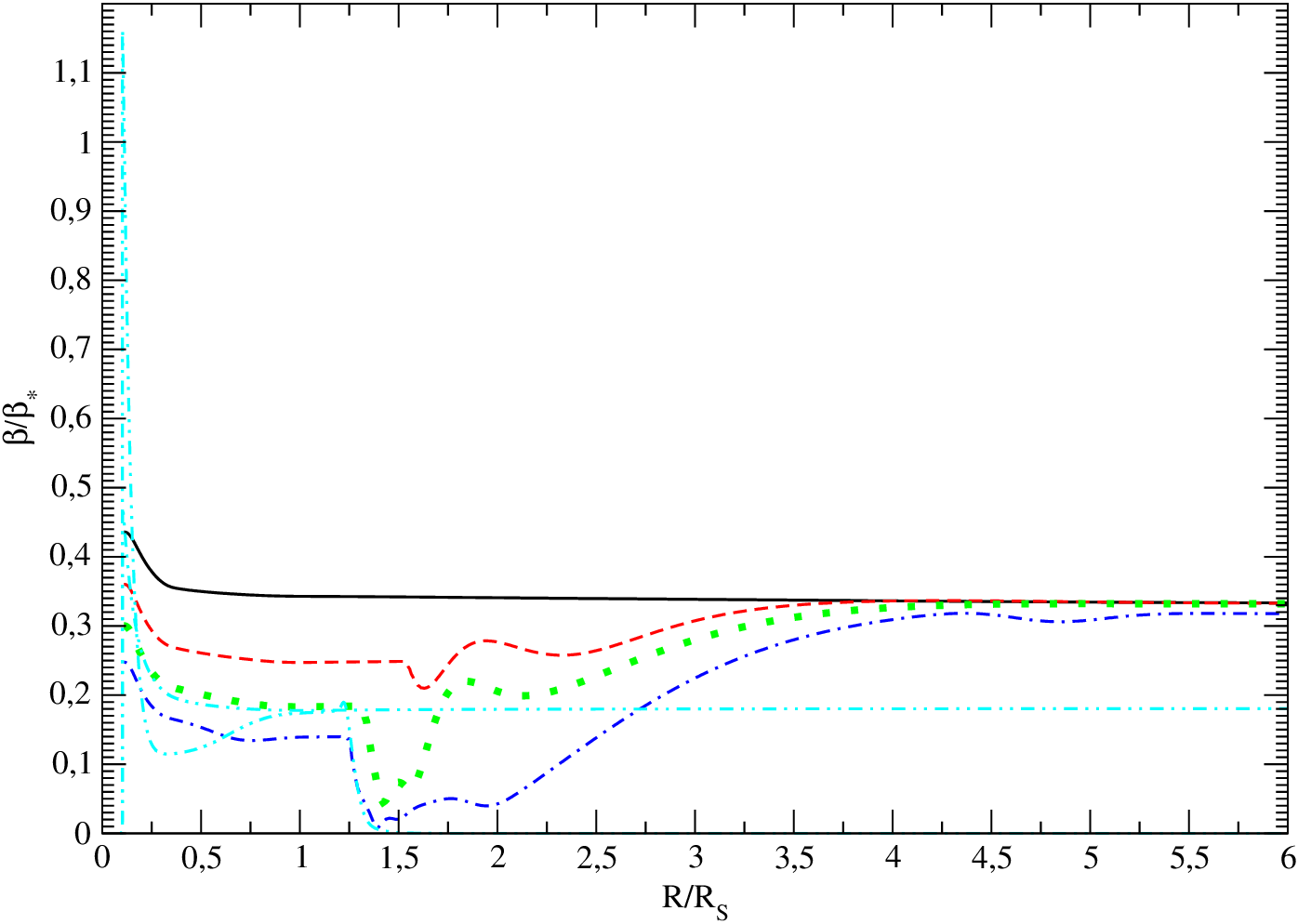}
\end{center}
\vspace{0cm}
\caption{Same as Fig. \ref{Fig19}, but for $\alpha=0.1$. Again, the dot dot dashed curve corresponding to $t=40P_{min}$ is
nearly flat at sufficiently large radii.}
\label{Fig21}
\vspace{-0.5cm}
\end{figure}

\begin{figure}
\begin{center}
\vspace{10cm}\hspace{-10cm}\includegraphics{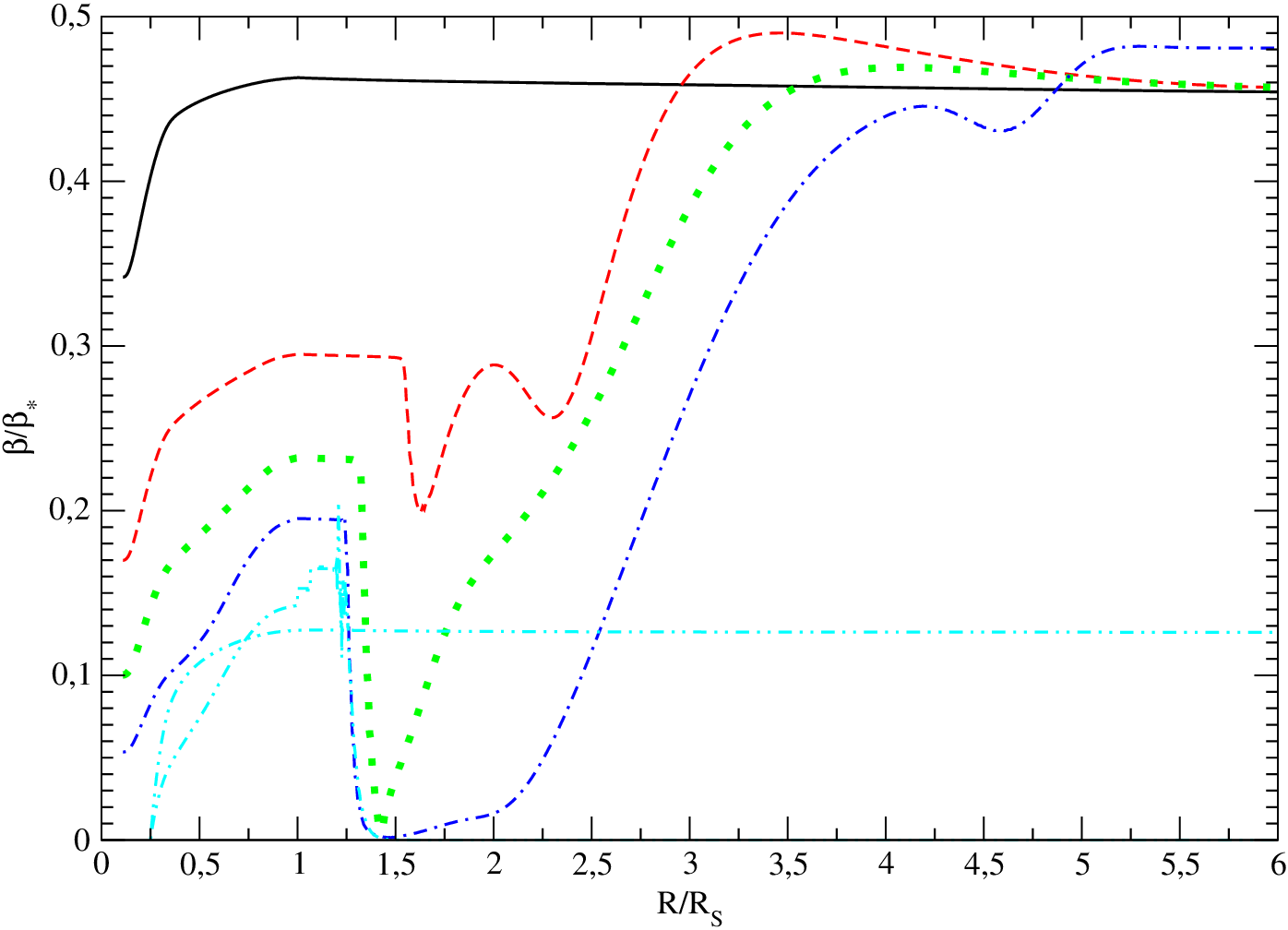}
\end{center}
\vspace{0cm}
\caption{Same as Fig. \ref{Fig21}, but for $a=-1$. 
Again, the nearly flat dot dot dashed curve is calculated in the framework of 
the quasi-static approach at time $t=40P_{min}$.}
\label{Fig22}
\vspace{-0.5cm}
\end{figure} 

Fig. \ref{Fig19a} shows profiles of $\beta$ calculated for the 'early' moments of time $t/P_{min}=11,12,13,14,15$ 
for the  case with $\alpha=0.1,$  $a=1,$ and  $t_{in}=10P_{min}$. It shows that similar to the case with $\alpha=0.01$  shown in Fig. \ref{Fig18}
the profiles are nearly flat, but, unlike that case there are no spikes in the  distributions of $\beta$. Indeed, as seen 
from  Fig. \ref{Fig5} the profiles of $\delta $ are rather uniform at these  moments of time for $\alpha=0.1$, so the 
conditions leading to the  formation  of a spike are not satisfied.

Figs \ref{Fig21} and \ref{Fig22} show the profiles of $\beta$ calculated at  the 'late' times for the case with $\alpha=0.1$,
for prograde and retrograde black hole rotations, respectively.
 As in the previous case we add two dot dot dashed curves 
to represent  results based on the quasi-stationary approximation for $t/P_{min}=40$ and $100$. 
Unlike the  previous case, 
even when $t/P_{min}=100$ the quasi-stationary curves are close to the fully numerical ones only at $R/R_{S} \sim 1$. 
At larger radii inclinations corresponding to the fully numerical calculations are significantly larger.
 
\noindent That difference can be explained by the fact that at larger radii a typical relaxation time to a quasi-stationary
configuration, $t_{\nu}\sim (\alpha / \delta^2)\Omega^{-1}$ (see \cite{Pap1983}) is larger than 
the time elapsed from the beginning of calculations. Indeed, using equation (\ref{eq3}) we can express $t_{\nu}$ 
as $5\cdot 10^{-3}(\alpha / \delta^{2})B_p^3B_S^{-3/2}P_{min}$. 
Substituting $\alpha=0.1$, $\delta \sim 10^{-3}$,
$B_p\approx 1.55$ and $B_S\approx 0.78$ we get $t_{\nu} \sim 200(R/R_S)^{3/2}P_{min}$, which indicates  that 
when $t \sim 100 P_{min}$ the disc is expected to be far from its stationary state.  However, we have checked  that when the computational domain s expanded outwards
using values of $\delta $ and $\Sigma $ at the outer boundary of the  computational grid used to calculate the background quantities,
inclinations at large radii drop significantly. Therefore, this effect seems to be sensitive to the size of computational domain and how the outer boundary condition was applied.

\subsection{Evolution with time of both inclination and rotation angle}\label{incnodes}

In order to graphically represent time evolution of both Euler angles $\beta$ and $\gamma $ characterising the position of the disc ring
with $R=R_S$ it is convenient to show time dependencies of  $W_1=Re({\cx W})=\beta \cos \gamma $ and $W_2=Im({\cx W})=\beta \sin 
\gamma$ evaluated at $R_S$. One can see that $W_1$ and $W_2$ represent represent a  vector perpendicular to the ring 
angular momentum vector such that its $x$ component is equal to $W_2$ and its $y$ component is equal to $-W_1$. In addition, they are proportional to the components of a vector lying along the line of intersection
of the plane of  a local disc ring and the equatorial plane (line of nodes).  Thus the time dependence of these 
components  provide information about both the  change of inclination and precession 
of this ring. A uniform precession  at constant inclination  corresponds to a sinusoidal oscillation. 
We show $W_1$ and $W_2$ as functions of time in Figs.
\ref{Fig23} and \ref{Fig24} respectively for $\alpha=0.01$ and $0.1.$ These were standard cases with $a=1$ and 
$t_{in}=10P_{min}$.

\begin{figure}
\begin{center}
\vspace{10cm}\hspace{-10cm}\includegraphics{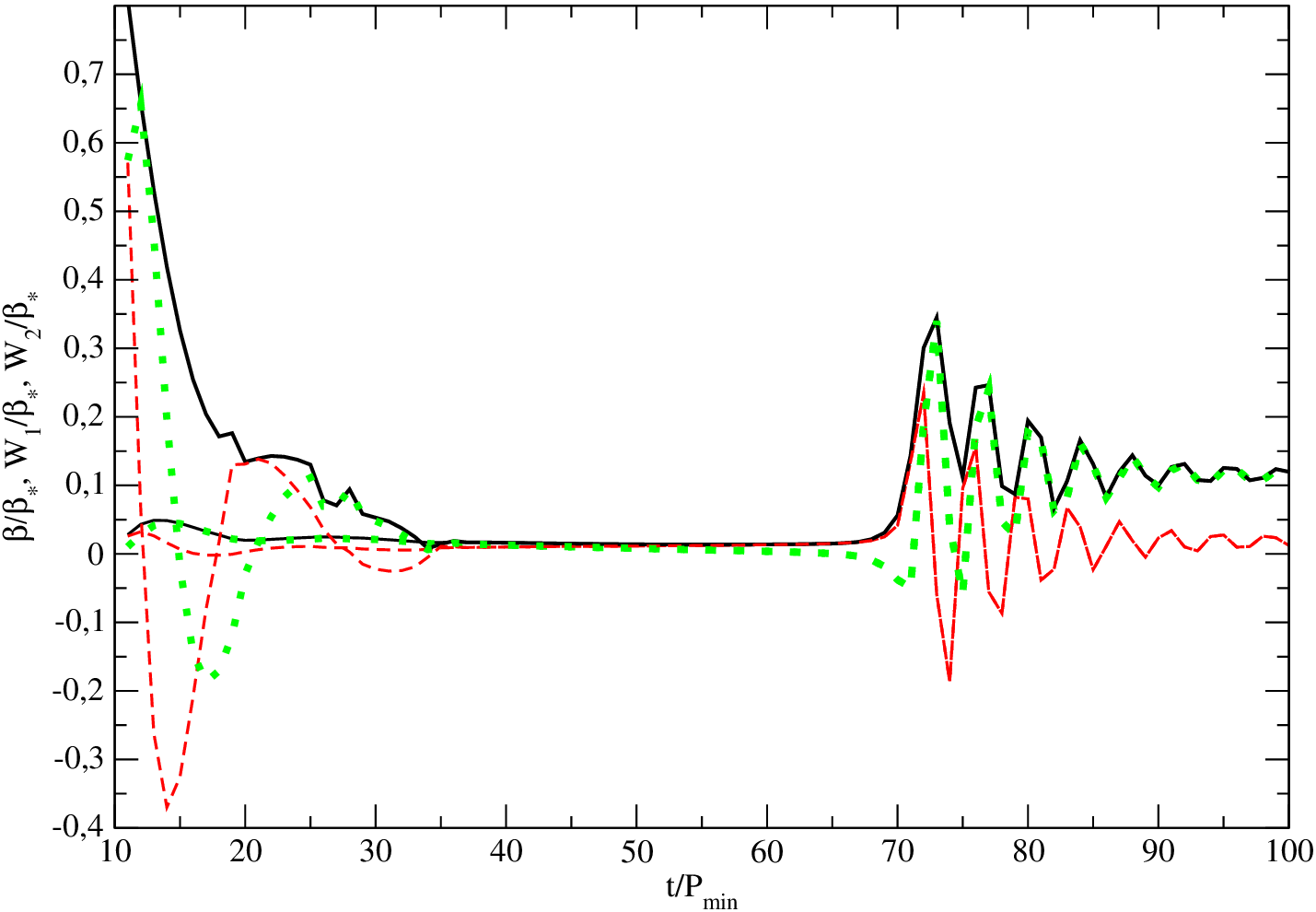}
\end{center}
\vspace{0cm}
\caption{Solid, dashed and dotted line show the evolution of $\beta$, $Re({\cx W})$, $Im({\cx W})$ evaluated 
at the stream impact position for our standard case
with $\alpha=0.01$, $a=1$ and $t_{in}=10P_{min}$. Curves having larger (smaller ) values of arguments 
at sufficiently small times correspond to the initial disc configuration being aligned with the stream orbital plane (black
hole equatorial plane). These curves become overlapping at large times.}
\label{Fig23}
\vspace{-0.5cm}
\end{figure}

\begin{figure}
\begin{center}
\vspace{10cm}\hspace{-10cm}\includegraphics{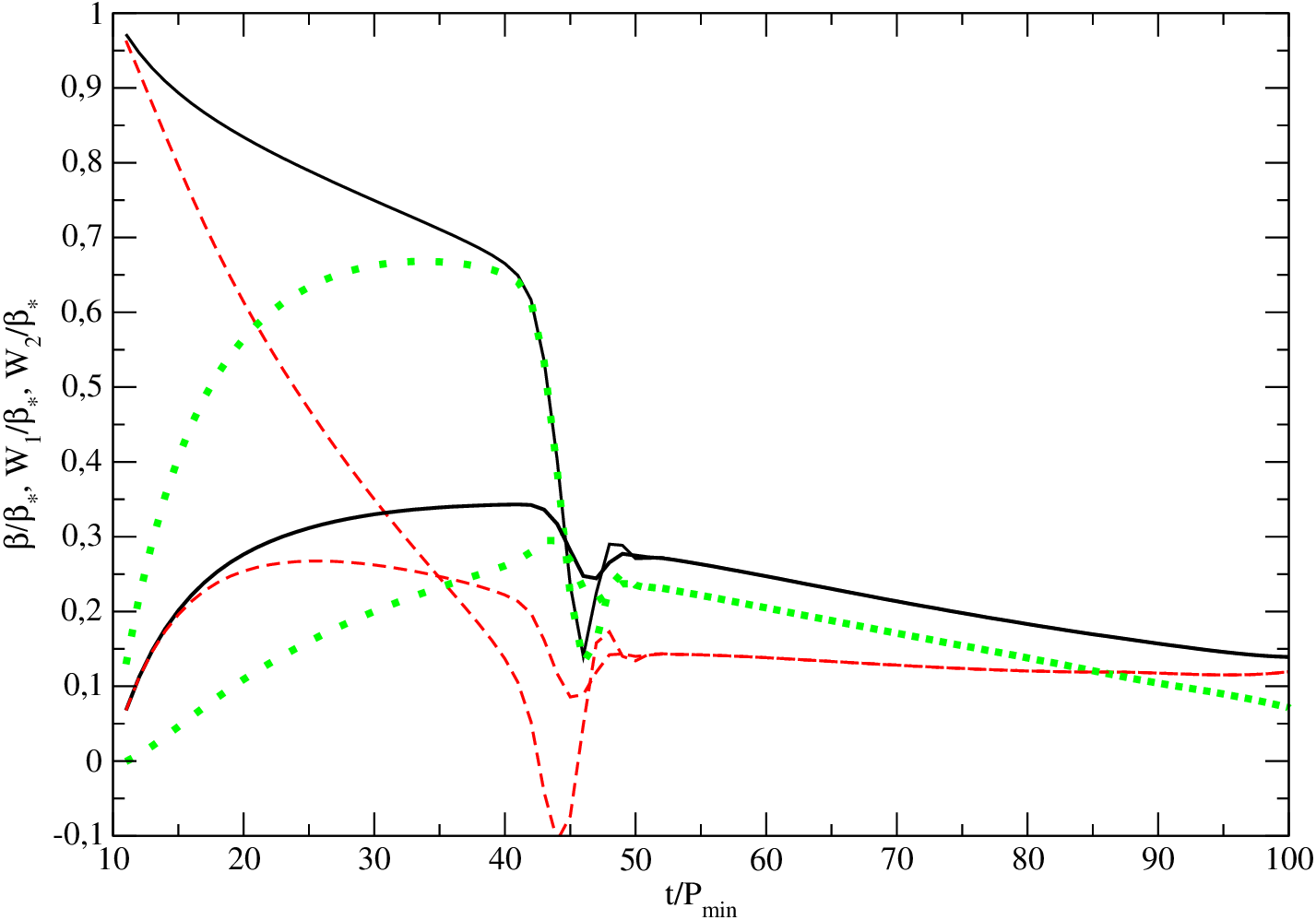}
\end{center}
\vspace{0cm}
\caption{Same as Fig. \ref{Fig23}, but for $\alpha=0.1$.}
\label{Fig24}
\vspace{-0.5cm}
\end{figure}

As seen from these Figures, during the early stage of evolution, when distributions of $\beta $ with $R$ are nearly flat
the ring accomplishes  only one precession period in the case of $\alpha=0.01$ and only a quarter of that  in case of $\alpha=0.1$.
Precession is also accompanied by a strong evolution of the ring's inclination due to the influence of the stream. That means 
that  for our parameters, the purely precessional model of disc  evolution considered in \cite{Sto2012} and \cite{Fra2016} is not applicable. Note, however, that this model could be more appropriate for  smaller values 
of $R_S$ leading to a stronger Lense-Thirring precession.


\section{Summary and conclusions}\label{Discuss}
\label{Disc}

In this Paper we consider the evolution of geometrical form of a disc formed after a tidal disruption event 
due to impact of material of the stream of gas, which arises from tidally disrupted star. Angular momentum 
provided by the stream is not, in general, aligned with angular momentum of black hole, so the stream tends
to 'push' the disc away from equatorial way, naturally leading to its twisted geometrical configuration.

We solve time dependent twisted disc equations numerically, for two values of viscosity parameter $\alpha=0.01$
and $0.1$, black hole mass $10^{6}M_{\odot}$ and various values of its rotational parameter $a$ in the linear regime. We find that at
relatively late times $t ~ > 50P_{min},$ with  $P_{min}$  being
 the minimum return time of gas in the stream to periastron after the disruption of the star,
 configurations of the twisted disc are close to those obtained in 
framework of a quasi-stationary approach  in which the background quantities are held constant in time 
while the inclination and twist are allowed to attain a steady state. This approach was adopted in
 \cite{Xiang2016}. 
 
 \subsection{Evolution of the twisted disc after tidal disruption}
However, at 'early' times $t ~ < 30P_{min}$ we found that
disc  shape is far from being quasi-stationary. At these times, inner parts of the disc are geometrically
rather thick, with relative thickness $\delta $ exceeding viscosity parameter $\alpha $. As was suggested elsewhere
\citep[see, e.g.][]{Sto2012} this leads to disc's inclination angle $\beta$ being nearly independent of radius 
in this region. However, unlike previous studies in Section \ref{incnodes}   we find that the evolution of disc's geometrical shape 
is not simply precessional, since it is determined by both Lense-Thirring torque and the torque arising from the
stream. Both inclination angle $\beta $ and precession angle $\gamma $ evolve on a similar time scale and for our simulations
with stream impact radius $R_S$ equal to  $9R_T/7,$ with $R_T$ being the tidal disruption radius, 
it took less than one precessional period for this stage to be completed. Note, however, that this would be
different for different parameters of the problem, for example  reducing $R_S$  would make the
Lense-Thirring torque stronger, thus speeding up precession.
 We propose a simple semi-analytic 
model for this stage in Section \ref{ODE} ( and see equation (\ref{e6})) , which is based on law of conservation of angular momentum and can be used without 
having to solve the twisted disc equations. It is enough to know  the  evolution of the  background
quantities $\delta$ and $\Sigma$ with time and
it allows inclination changes and the amount of precession to be estimated.
The duration of this stage is determined by the time needed for the disc
to cool down significantly at $R\sim R_S$ at which point $\delta $  becomes small on this scale.
\subsection{Outward propagating transition front}
For  the case with  $\alpha=0.01,$  in Section \ref{propfront} we found that during the early stage of evolution,  the disc exhibits  singular
behaviour of its inclination angle.
There is an outward propagating  transition front that separates an inner hot region from an outer cooler region
 which constitutes a pre-existing low surface density  disc that is coplanar with the black hole equatorial plane.
The transition  radius $R_{crit}$  eventually extends to $\sim 13R_{S}$. At radii smaller than $R_{crit}$
$\beta $ is nearly uniform, at large radii it is close to zero, while in a narrow region close to $R_{crit}$ 
there is a sharp growth of $\beta $ leading to formation of a spike in its profile.
 
 When the  
 numerical scheme  is not regularised formation of this spike eventually leads to a numerical instability.
  We regularised 
our equations by  adding additional  dissipation in this region parametrising by a value of the coefficient
$\alpha_1$ at the transition front  which is chosen to be $1000\alpha$  for most simulations.
 We checked that  the enhancement of 
$\alpha_1$  changed only the spike amplitude significantly,   leaving  the solutions outside the  region of the spike
practically unchanged. Physically, formation of this spike is related to the  outward propagation of  transition front 
that separates hot and cold regions.
 In a low viscosity disc tilt and twist propagate as a  bending wave with  speed equal
to half the sound speed. At the point where the speed of bending waves becomes equal to the  propagation 
speed of the front  wave action  accumulates  leading to singularity and numerical instability. 
In Section \ref{simplefront}  we formulated a simple
analytic theory of distribution of $\beta $ in this region and checked that it is confirmed by numerical simulations.

A question arises as to what is the physical mechanism that limits the spike amplitude. One  possibility is 
the action of non-linear effects. In this case it's possible to speculate that the spike's amplitude could
attain values of order unity,  $\beta \sim 1$, although various instabilities \citep[see e.g.][] {Fer2008, Fer2009, Lat2013a, Lat2013b}
 could significantly limit its amplitude.  If the spike has large amplitude, the disc may take on a broken structure
of the type postulated by e.g. \citet{Nea2016},
  which could produce  significant observational effects through intercepting radiation from the central source. 
Another possibility would be to equations for the disc tilt and inclination  to  consider terms of higher  order in 
in $\delta$.  Both these possibilities deserve a future study.   
   
   \subsection{Inclination of the disc at the stream impact radius}
In Section \ref{streaminc}  we calculated the  time dependence of the  disc inclination at the stream impact radius for different $\alpha $, $a$ and
initial conditions. Similar to  what was found using the quasi-stationary approach we found  that the disc inclination could be large, 
being of the order $0.1-0.5\beta_*$, where $\beta_*$ is  the inclination of the  orbital plane of the stream, 
for $|a|=1$. For smaller black hole
rotation the inclination angle is even larger. The inclination angle attains
its largest values as a function of time during the transition from an  advective to radiative phase
occurring when  $t\sim 50-70P_{min}$. Therefore, observations of TDEs during this time could shed light on physical conditions during
this transition,  including whether  the thermal instability operates during this time.    

\subsection{Discussion}
That the disc inclination angle could be large and is a non-trivial function of time and radius could have profound 
implications on time behaviour of disc's luminosity and other effects associated with TDEs. 
 Clearly, disc's luminosity will be modulated by  the changing  geometrical shape, which would allow us, 
 in, principal, to test different models of the  accretion  process and provide information on black hole mass and angular momentum. 
 Moreover, it was suggested recently that  when  a sufficiently thick disc  is inclined with respect to
 the  equatorial plane as expected at the 'early'  evolution times, it  produces a jet directed perpendicular to the disc plane, see \cite{Lis2018}.
  Therefore, the evolution of the  geometrical form of the disc may also lead to evolution of the  jet  luminosity.
  The issue of  the observational appearance of such discs is to be considered in future publications.

\section*{Acknowledgements}

PBI was supported in part by RFBR grants 16-02-01043 and 17-52-45053 and in part by Programme 28 of the Fundamental Research of the Presidium of the RAS, VVZ was supported by grant RSF 14-12-00146 for obtaining the numerical solutions of governing equations for twisted disc.


\begin{appendix}

\section{An outline of the derivation of the governing equation for twisted disc dynamics}
\label{outline}
In order to derive the relativistic twist equations appropriate for our model, we start from  equations (46-48) of
\cite{Zhu2011}   and consider the case of  a vertically isothermal disc with gas density depending on vertical coordinate 
according to equation (\ref{en1}) where $z$ is understood as a proper distance from the equatorial plane of the twisted disc.

Using their equations (46) and (47) one  may derive an equation describing  the evolution of the radial velocity perturbation, $v^r$, induced by the disc twist and warp.
In disc with  density distribution given by (\ref{en1}) $v^r$ can be represented by the complex state variable ${\cx B} = B_2 + {\rm i}B_1$, which
we introduce here using the following representation
\begin{equation}
\label{def_B}
v^r = z (B_1 \sin\varphi + B_2 \cos\varphi).
\end{equation}

The derivation of the equation for ${\cx B}$ follows the same procedure as that given in Section 3.3 of \cite{Zhu2011}, see their final equation (60), with a remark that 
here we use the Schwarzschild radial coordinate and the proper vertical coordinate    rather than the 
so-called isotropic radial coordinate, $R_{iso}$, and a similar vertical coordinate, related to $R$ and $z$ through
\begin{equation}
\label{R_xi}
R = K_2 R_{iso}, \quad z = K_2 z_{iso},
\end{equation}
where $K_2 = (1+1/(2R_{iso}))^2$   that was used by  \cite{Zhu2011} .
Taking this into account, as well as the expression for $r\varphi-$component of the stress tensor of the background flow
\begin{equation}
\label{T_rphi}
T^{r\varphi} = \frac{3}{2} \nu \rho K_1 (U^\tau)^2 \frac{U^\varphi}{R},
\end{equation}
where $\nu$ is given by equation (\ref{en1a}),
we obtain a equation (\ref{eq_B}).

The dynamical equation (\ref{eqW})  for ${\cx W}$  follows from equation (48) of \cite{Zhu2011}  after making use of
 the relations (\ref{R_xi}), the definition of $\Sigma$ as well as the density  distribution (\ref{en1}) together with
the explicit form of $T^{r\varphi}$, where $\nu$ is given by equation (\ref{en1a}). Note that $\Sigma = K_2 \Sigma_{ZI}$, where $\Sigma_{ZI}$ is the surface density  used
in equation (48) of \cite{Zhu2011}.

\section{Details of the  numerical  procedure for solving  the governing equations }

\subsection{Extrapolation of dynamical variables to smaller  radii}
\label{extension}

Our inner boundary conditions are set at the radius of the  marginally stable orbit of a  non-rotating black hole,
$R_{ms}=6GM/c^2$, where our  equations  governing the dynamics are formally singular. From these it follows that
in order to have regular solutions we must require $\partial {\cx W}/ \partial R={\cx B}=0$ at that location. 
In order to extend the computational domain  used to calculate the background quantities  to  include the marginally stable orbit
we need to make some assumptions about the  behaviour of the state variables at small radii. 
We assume that $\Sigma$ and $\delta $
correspond to a Novikov-Thorne solution near $R=R_{ms}$, which is smoothly matched to the  exterior numerical solution 
through an intermediate matching region. Details of our matching procedure are given below.

Let $R_i$ be the inner boundary of the background solution obtained numerically using the Newtonian numerical model, 
which prescribes values of $\delta_{num}$ and $\Sigma_{num}$ as functions of $R$ and $t.$ 
As discussed in the main text we multiply $\delta_{num}$ and $\Sigma_{num}$ by factors $D(R)^{3/20}$ and $D(R)^{7/10}$,
respectively, where the function $D(R)$ is defined through  equation (\ref{eq2011c}) to obtain background 
quantities $\delta(t, R)$ and $\Sigma (t,R)$ that  are used in equations (\ref{eq_B}) and (\ref{eqW}). 
This procedure is needed  in order to match  to a quasi-stationary, formally thin, relativistic disc, which must have a fixed value of specific angular momentum at $R=R_{ms}.$ 

We assume that starting from some $R_{NT} <R_i$ the disc is represented by a Novikov-Thorne analytic solution with $\delta=\delta_{NT}D^{3/20}$, where $\delta_{NT}$ is a constant. 
Further, in an intermediate domain $R_{NT} < R < R_i$ we specify
\begin{equation}
\label{delta_im}
\delta_{im} = \delta_i + \delta^\prime_i (R-R_i) + C_\delta (R-R_i)^2
\end{equation}
 where $\delta_{im}$ is the aspect ratio  in the intermediate domain and
\begin{equation}
\label{Sigma_im}
\Sigma_{im} = \Sigma_i + \Sigma^\prime_i (R-R_i) + C_\Sigma (R-R_i)^2
\end {equation}
where  $\Sigma_{im}$ is the  surface density there.
 The quantities $\delta^\prime_i$  and  $\Sigma^\prime_i$ are, respectively, the derivatives of 
 $\delta_{im}$ and  $\Sigma_{im}$
 found  from the numerical solution at $R=R_i$. 

The unknown coefficients $C_\delta$ and $C_\Sigma$ are determined by the requirement that the intermediate solution 
specified by (\ref{delta_im}-\ref{Sigma_im})  matches smoothly to the Novikov-Thorne solution at $R=R_{NT}.$
We define $\Delta \equiv R_i - R_{NT}$.
In order to match smoothly at $R=R_{NT}$ we require that
$$
\frac{d\delta_{im}}{dR}\Biggr |_{R=R_{i}} = \frac{d\delta_{NT}}{dR}\Biggr |_{R=R_{NT}} =  0,
$$
This requires that  that $C_\delta=\delta^\prime_i/(2\Delta)$. 
Consequently we have
$$
\delta_{NT} = \delta_i + \delta^\prime_i (R_{NT}-R_i) + \frac{\delta^\prime_i}{2\Delta} (R_{NT}-R_i)^2.
$$

Under the assumption that the turbulent kinematic viscosity coefficient is parametrised via $\alpha$ as
given by equation (\ref{en1a}),
the Novikov-Thorne solution yields the following form of surface density  \citep [see equation (36) of][] {Zhu2011}
\begin{equation}
\label{Sigma_NT}
\Sigma_{NT} = \frac{\dot M}{3\pi\alpha} \frac{D(R)}{K_1^3 U^\tau (U^\varphi)^2 R^{3/2} \delta_{NT}^2},
\end{equation}
where $U^{\tau}$ and $U^{\varphi}$ are given by equation (\ref{eq2011b}) and $D(R)$ is given by equation 
(\ref{eq2011c})
\newline and $K_1 = \sqrt{1 - 2/R}.$
Note that equation (\ref{Sigma_NT}) may be expressed as
\begin{equation}
\label{F}
\Sigma_{NT} = \frac{\dot M}{\delta_{NT}^2} F(R),
\end{equation}
where $F(R)$ is a known function of $R$.

Matching  $\Sigma$ given by equation (\ref{Sigma_im}) in the intermediate domain to $\Sigma_{NT}$ at $R=R_{NT}$
requires that 
\begin{equation}
\label{Sig_match_1}
\Sigma_{im}(R_{NT})=\Sigma_{NT}(R_{NT})
\end{equation}
and matching derivatives requires that
\begin{equation}
\label{Sig_match_2}
\frac{d\Sigma_{im}}{dR}\Biggr |_{R=R_{NT}} = \frac{d\Sigma_{NT}}{dR}\Biggr |_{R=R_{NT}}
\end{equation}
These  conditions require that
\begin{equation}
\label{C_Sigma}
C_\Sigma = \frac{\Sigma^\prime_i/F^\prime  + \Delta \Sigma^\prime_i/F  - \Sigma_i/F }{\Delta^2/F + 2\Delta/F^\prime},
\end{equation}
where $F$ and its derivative, $F^\prime\equiv dF/dR$, are evaluated  at $R=R_{NT}$.

Thus, the disc aspect ratio and the surface density are given by the numerical solution for  $R>R_i$, 
by the intermediate solution (\ref{delta_im}-\ref{Sigma_im}) for  $R_i > R > R_{NT}$ and by the Novikov-Thorne solution 
(\ref{Sigma_NT}) with constant $\delta=\delta_{NT}$ for  $R<R_{NT}$.
For all computations we set $R_{NT}/R_i=0.85$ and it has been  checked that  solutions  of the twist equations are insensitive to this choice.

\subsection{An implicit grid based scheme}
\label{scheme}
In order to construct the numerical scheme we adopted,  let us rewrite equations (\ref{eq_B}) and (\ref{eqW}) in the form
\begin{equation}
\label{eq_B1}
\dot {\cx B} = {\rm i} {\cx O} {\cx B} + {\cx F} {\cx W}',
\end{equation}
\begin{equation}
\label{eq_W}
\dot {\cx W} = {\rm i} \Omega_{LT} {\cx W} + G {\cx W}' + C({\cx D} {\cx B} + E{\cx W}')' + \dot {\cx W}_S,
\end{equation}
\vspace{1cm}
where we represent the source term (\ref{eq11}) as a combination of two factors
$$
{\dot {\cx W}}_S \equiv {\tilde {\dot {\cx W}}}_S - {\bar {\dot W}}_S {\cx W},
$$
where ${\tilde {\dot {\cx W}}}_S$ and ${\bar {\dot W}}_S$ can be easily read off from (\ref{eq11}) and  
${\cx O}, {\cx F}, {\cx D}$ and $G, C, E$ are known, respectively, complex and real functions of time and radial coordinate, while the time and the spatial 
partial derivatives are denoted by dot and prime, respectively.

Let us set up a  uniform spatial grid in the variable $x \equiv \sqrt{R}$, i.e. $x_j = \sqrt{6}+ \delta x + j \Delta x$, 
where $j = 0, 1, 2, .. N$ specifies the spatial node number, $\Delta x$ is the step length and $\delta x$ is the grid offset which is used to avoid the  singularity occurring
in the coefficients of our equations at the last stable orbit occurring at  $x =\sqrt{R}=\sqrt{6}.$ We recall that  in setting up these coordinates
spatial and temporal scales are expressed in terms of
 $GM/c^2$ and $GM/c^3$, respectively. The total number of grid points is $N=1.$  
We denote the time step by $\Delta \tau$.  The $n_{th}$  time slice is defined to be  at time
 $\tau_{n} = \tau_{n-1} + \Delta \tau.$ 
The radial extent of the computational domain is $N \Delta x.$ 
It is assumed that the numerical approximations to the time and spatial derivative of some 
quantity $A$ are centered at the times  $\tau_{n+1/2} = (\tau_{n}+\tau_{n+1})/2$ and at the coordinates $x_{j}$ and are given by the following expressions:

\begin{equation}
\label{num_1}
(\dot A)^{n+1/2}_j = \frac{A^{n+1}_{j} - A^{n}_j}{\Delta \tau},
\end{equation}
\begin{equation}
\label{num_2}
(A^\prime)^{n+1/2}_{j} = \frac{A^{n+1}_{j+1} + A^{n}_{j+1} - A^{n+1}_{j-1} - A^{n}_{j-1}}{4\Delta x}.
\end{equation}
while the quantity itself is given by
\begin{equation}
\label{num_3}
A^{n+1/2}_j = \frac{A^{n+1}_{j} + A^{n}_j}{2}.
\end{equation}

By $A$ we indicate any of the variables entering the twist equations except 
the term $(E {\cx W}^\prime)^\prime$ which is approximated  according to
\begin{equation}
\label{num_4}
((E {\cx W}^\prime)^\prime) ^{n+1/2}_{j}= \frac{(E {\cx W}^\prime)^{n+1/2}_{j+1/2} - (E {\cx W}^\prime)^{n+1/2}_{j-1/2} }{\Delta x},
\end{equation}
where
\begin{equation}
\label{num_5}
(E {\cx W}^\prime)^{n+1/2}_{j+1/2} = E^{n+1/2}_{j+1/2} \, \frac{ {\cx W}^{n+1}_{j+1} + {\cx W}^{n}_{j+1} - {\cx W}^{n+1}_j - {\cx W}^n_j  }{2\Delta x}
\end{equation}
and
\begin{equation}
\label{num_6}
(E {\cx W}^\prime)^{n+1/2}_{j-1/2} = E^{n+1/2}_{j-1/2} \, \frac{ {\cx W}^{n+1}_{j} + {\cx W}^{n}_{j} - {\cx W}^{n+1}_{j-1} - {\cx W}^n_{j-1} }{2\Delta x}
\end{equation}
with
\begin{equation}
\label{num_7}
E^{n+1/2}_{j \pm 1/2} = \frac{E^{n+1/2}_j + E^{n+1/2}_{j \pm 1}}{2}, 
\end{equation}
From the numerical representation of (\ref{eq_B1})   and  (\ref{eq_W})
made with the help of (\ref{num_1}-\ref{num_7}) we obtain  $N-1$ linear inhomogeneous algebraic equations for the state vector 
\begin{equation}
\label{num_8}
{\bf q} \equiv [ {\cx B}^{n+1}_0,\,{\cx W}^{n+1}_0,\, ...,\,{\cx B}^{n+1}_j,\,{\cx W}^{n+1}_j,\, ...,\, {\cx B}^{n+1}_N,\,{\cx W}^{n+1}_N  ]^\dag
\end{equation}
relating the variables at the next slice, $n+1,$ to those  at the slice $n.$
These can be represented by the matrix equation
\begin{equation}
\label{num_9}
{\bf M } {\bf q} = {\bf B }.
\end{equation}
In equation (\ref{num_9}) the non-zero elements of ${\bf M}$ are given by
\begin{align}
\label{num_10}
&{\bf M}_{2j-1}^{2j} = \frac{\Delta\tau}{4\Delta x} {\cx F}^{n+1/2}_j, \nonumber \\
 &{\bf M}_{2j}^{2j}  = 1 - \frac{{\rm i}\Delta\tau}{2} {\cx O}^{n+1}_j, \nonumber \\
&{\bf M}_{2j+3}^{2j}  = - \frac{\Delta\tau}{4\Delta x} {\cx F}^{n+1/2}_j, \nonumber \\
&{\bf M}_{2j-2}^{2j+1}  = \frac{\Delta\tau}{4\Delta x} C^{n+1/2}_j {\cx D}^{n+1}_{j-1}, \nonumber \\
&{\bf M}_{2j-1}^{2j+1} = \frac{\Delta\tau}{4\Delta x} G^{n+1/2}_j - \frac{\Delta\tau}{2\Delta x^2} C^{n+1/2}_j E^{n+1/2}_{j-1/2}, \\
&{\bf M}_{2j+1}^{2j+1} = 1 - \frac{ {\rm i} \Delta\tau}{2} (\Omega_{LT})^{n+1}_j + \frac{\Delta \tau}{2\Delta x^2} C^{n+1/2}_j (E^{n+1/2}_{j+1/2} + E^{n+1/2}_{j-1/2})
+ \frac{ ( {\bar {\dot W}}_S )^{n+1}_j }{2}\Delta \tau, \nonumber \\
&{\bf M}_{2j+2}^{2j+1} = - \frac{\Delta\tau}{4\Delta x} C^{n+1/2}_j {\cx D}^{n+1}_{j+1}, \nonumber \\
&{\bf M}_{2j+3}^{2j+1} = - \frac{\Delta\tau}{4\Delta x} G^{n+1/2}_j - \frac{\Delta\tau}{2\Delta x^2} C^{n+1/2}_j E^{n+1/2}_{j+1/2}, \nonumber
\end{align}
while the non-zero elements of ${\bf B}$ are explicitly 
\begin{align}
\label{num_11}
&{\bf B}^{2j} = {\cx B}^n_j \left ( 1 + \frac{{\rm i} \Delta \tau}{2} \Omega^n_j \right ) + \frac{\Delta t}{4\Delta x} F^{n+1/2}_j ({\cx W}^n_{j+1}-{\cx W}^n_{j-1}), \nonumber \\
&{\bf B}^{2j+1} = {\cx W}^n_j \left ( 1 + \frac{{\rm i} \Delta \tau}{2} (\Omega_{LT})^n_j \right ) + \frac{\Delta \tau}{4\Delta x} G^{n+1/2}_{j} ({\cx W}^n_{j+1} - 
{\cx W}^n_{j-1}) +\nonumber \\
&\frac{\Delta \tau}{4\Delta x} C^{n+1/2}_j ( {\cx D}^n_{j+1} {\cx B}^n_{j+1} - {\cx D}^n_{j-1} {\cx B}^n_{j-1} ) +  
\frac{\Delta \tau}{2\Delta x^2} C^{n+1/2}_j ( E^{n+1/2}_{j+1/2} ({\cx W}^{n}_{j+1}-{\cx W}^n_j)  \nonumber \\
&- E^{n+1/2}_{j-1/2} ({\cx W}^n_{j} - {\cx W}^n_{j-1}) ) 
+ ( {\tilde {\dot {\cx W}}}_S )^{n+1/2}_j \Delta \tau - \frac{ ({\bar {\dot W}}_S )^{n}_j {\cx W}^n_j }{2} \Delta \tau 
\end{align}
In the expressions (\ref{num_10}) and (\ref{num_11}) it is implied that the subscripts and the superscripts of elements
of  ${\bf M}$ and ${\bf B}$ denote the columns and the row  numbers, respectively, 
with  $j$ taking on  values from $1$ to $N-1$.

Equation (\ref{num_9}) must be completed by 
representations of the boundary conditions.
The latter are the same as in \cite{Zhu2014}, explicitly
\begin{equation}
\label{num_12}
{\cx W}^{\prime} = 0\quad{\rm and}\quad 
{\cx B} = 0
\end{equation}
at $x = \sqrt{6} +\delta x$ and $x = \sqrt{6} +\delta x + N\Delta x$.
This yields the additional conditions
$$
{\cx W}^{n+1}_0 = {\cx W}^{n+1}_1, \quad {\cx B}^{n+1}_0 = 0
\quad {\rm together  \quad  with }\hspace{2mm}
{\cx W}^{n+1}_N = {\cx W}^{n+1}_{N-1}  \quad {\rm and} \quad {\cx B}^{n+1}_N = 0.
$$
Incorporating  these conditions in the system  (\ref{num_9})
leads to additional non-zero elements of ${\bf M}$:
\begin{eqnarray}
\label{num_14}
{\bf M}_0^0 = 1, \quad
{\bf M}_1^1 = 1, \quad
{\bf M}_3^1 = -1, \quad
{\bf M}_{2N}^{2N} = 1, \quad   
{\bf M}_{2N-1}^{2N+1} = -1 \quad {\rm and}\quad
{\bf M}_{2N+1}^{2N+1} = 1.
\end{eqnarray}

\noindent The system is  solved numerically using the Gaussian elimination method adapted for almost diagonal matrices.
In addition the convergence of the solution is  checked  by making sure that an angular momentum conservation law, which is derived from equation (\ref{eq_W}) multiplying it
by $C^{-1}$ and subsequently  integrating over the spatial domain, is satisfied.

\end{appendix}

\label{lastpage}

\end{document}